\documentclass[aps,superscriptaddress, showpacs,preprintnumbers, superscriptaddress, nofootinbibt,twocolumn]{revtex4}
\usepackage{amssymb}

\def\be{\begin{equation}}
\def\ee{\end{equation}}
\def\bea{\begin{eqnarray}}
\def\eea{\end{eqnarray}}

\begin{document}

\title{Quantum Cosmology of $f(R,T)$ gravity}

\author{Min-Xing Xu}
\email{253659701@qq.com}
\affiliation{School of Physics, Sun Yat-Sen University, Guangzhou 510275, P. R. China}
\affiliation{Yat Sen School, Sun Yat-Sen University, Guangzhou 510275, P. R. China}
\author{Tiberiu Harko}
\email{t.harko@ucl.ac.uk}
\affiliation{Department of Physics, Babes-Bolyai University, Kogalniceanu Street, Cluj-Napoca 400084, Romania}
\affiliation{Department of Mathematics, University College London, Gower Street, London
WC1E 6BT, United Kingdom}
\author{Shi-Dong Liang}
\email{stslsd@mail.sysu.edu.cn}
\affiliation{School of Physics, Sun Yat-Sen University, Guangzhou 510275, P. R. China}
\affiliation{State Key Laboratory of Optoelectronic Material and Technology, and
Guangdong Province Key Laboratory of Display Material and Technology, Guangzhou, P. R. China}

\date{\today }

\begin{abstract}
Modified gravity theories have the potential of explaining the recent acceleration of the Universe without resorting to the mysterious concept of dark energy. In particular, it has been pointed out that matter-geometry coupling may be responsible for the recent cosmological dynamics of the Universe, and matter itself may play a more fundamental role in the description of the gravitational processes that usually assumed. In the present paper we study the quantum cosmology of the $f(R,T)$ gravity theory, in which the effective Lagrangian of the gravitational field is given by an arbitrary function of the Ricci scalar, and the trace of the matter energy-momentum tensor, respectively. For the background geometry we adopt the Friedmann--Robertson--Walker metric, and we assume that matter content of the Universe consists of a perfect fluid.  In this framework we obtain the general form of the gravitational Hamiltonian, of the quantum potential, and of the canonical momenta, respectively.  This allows us to formulate the full Wheeler-de Witt equation describing the quantum properties of this modified gravity model. As a specific application we consider in detail the quantum cosmology of the $f(R,T)=F^0(R)+\theta RT$ model, in which $F^0(R)$ is an arbitrary function of the Ricci scalar, and $\theta $ is a function of the scale factor only. The Hamiltonian form of the equations of motion, and the Wheeler-de Witt equations are obtained, and a time parameter for the corresponding dynamical system is identified, which allows to formulate
the Schr\"{o}dinger--Wheeler--de Witt equation for the quantum-mechanical description of
the model under consideration. A perturbative approach for the study of this equation is developed, and the energy levels of the Universe are obtained by using a twofold degenerate perturbation approach. A second quantization approach for the description of quantum time is also proposed, and briefly discussed.

\pacs{04.20.Cv; 04.50.Kd; 04.60.Bc; 04.60.Ds}

\end{abstract}

\maketitle

\section{Introduction}

 One of the cornerstones of theoretical physics, General Relativity (GR), formulated mathematically in terms of the Einstein field equations, proved
to be a very successful gravitational theory at the scale of the Solar
System. By using GR we can describe the gravitational dynamics of the Solar system with a high precision, and phenomena like the  Mercury perihelion precession, the bending of light while passing the Sun, and the gravitational redshift can be fully understood.  In GR the gravitational field equations can be obtained by varying the Einstein-Hilbert action $S = \frac{1}{16\pi}\int{ R \sqrt{-g}d^4x} +\int {L_m \sqrt{-g}d^4 x}$, where $R$ is the Ricci scalar, and $L_m$ is the matter Lagrangian,  with respect to the metric $ g^{\mu\nu}$, and they read as $R_{\mu\nu} -\frac{1}{2}g_{\mu\nu} R =8\pi T_{\mu\nu}$, where $T_{\mu \nu}$ is the energy-momentum tensor. The energy-momentum tensor $T_{\mu \nu}$  identically satisfies the mathematical relation $\nabla ^{\mu} T_{\mu\nu} =0$, which can be interpreted from the physical point of view as the energy conservation. Essentially GR is a beautiful geometric theory that establishes a deep connection between the geometry of the spacetime, matter fields, and gravitational interaction. Considering larger scales of the Universe, using GR we can numerically simulate galaxies' formations and collisions, and the results of these simulations can be verified by the increasingly trustworthy data obtained due to the rapidly improving observational techniques. The predictions of general relativity have also been confirmed in the strong
gravity regime by the discovery of the gravitational wave emission in the binary
pulsars system PSR 1913 + 16 \cite{Hulse}, a discovery that has opened a new testing ground
for GR and for its generalizations. The detection by LIGO of GW150914 from the inspiral and
merger of a pair of black holes \cite{Abbott} will start a new era
in observational  astronomy, based on the theoretical and mathematical formalism of GR.

Thus, with a large number of astronomical observations, as well as terrestrial
experiments testing and confirming it both at the weak and strong
gravity regime, it would seem that GR gives a full description of the gravitational interaction at the non-quantum level. However, several recent astrophysical and cosmological observations have raised the intriguing possibility that GR may not be able to model and explain the gravitational dynamics
at scales much larger that the one of the Solar System.

On a fundamental theoretical level the two most important challenges GR must face  are the dark energy and the dark matter problems, respectively.Several cosmological observations, obtained initially from the study of the distant
Type Ia Supernovae, have provided the unexpected result that the expansion of the Universe did accelerate
lately \cite{1n,2n,3n,4n,acc}. The paradigmatic and usual explanation of the late time acceleration requires the existence of a mysterious and dominant component of the Universe, called
dark energy (DE). Dark energy  is responsible for the
late time dynamics of the Universe \cite{PeRa03,Pa03}, and can explain the observed features of the recent cosmological evolution. The second and equally mysterious component in the Universe, called Dark Matter, an assumed non-baryonic and non-relativistic "substance",  is necessary for the explanation of the flat rotation curves of galaxies, and for
the virial mass discrepancy in clusters of galaxies \cite{dm1,dm2}. The detection/observation of dark matter is  restricted by the fact that it interacts only gravitationally. Its effects can be observed
by observations of the motion of the massive hydrogen clouds around galaxies, or by the motion of the galaxies in clusters \cite{dm1}. However, despite many decades of intensive observational and experimental efforts the particle nature
of dark matter still remains essentially unknown.

One interesting possibility for explaining dark energy is based on theoretical models that
contain a mixture of cold dark matter and a
slowly-varying, spatially inhomogeneous component, called quintessence \cite{8n}. The idea of
quintessence can be implemented theoretically by assuming that it is the energy associated
with a scalar field $Q$, having a self-interaction potential $V(Q)$, and a pressure $p=\dot{Q}%
^{2}/2-V(Q) $ associated to the quintessence $Q$-field. Such a model also allows  a possible
theoretical interpretation in terms of particle physics results. If
the potential energy density $V(Q)$ of the quintessence field is much greater
than the kinetic one, then it follows that the pressure $p$ of the field is negative. Quintessential cosmological models have been extensively
investigated in the physical literature (for a recent review of quintessence cosmologies see \cite{Tsu}).

Alternatively,  the recent acceleration of the Universe can also be explained by scalar fields $\phi $ that are minimally coupled to gravity via a negative
kinetic energy, known as phantom fields, which  have been proposed in \cite{phan1}. Interestingly enough, these cosmological models  allow values of the parameter of the equation of state $w$ with $w<-1$.  Hence, real or complex scalar fields may play a fundamental role in the cosmological processes describing the evolution of our Universe, and they may provide a realistic description of the observed cosmic dynamics.

However, in order to explain the observed gravitational dynamics of the Universe a different line of thought on dark energy was also considered. It is based on the fundamental idea that dark energy is not a particular physical field, but it can be understood as a gravitational phenomenon induced at cosmological scales by the {\it intrinsic modifications of the gravitational interaction itself}. Hence the fundamental assumption of this theoretical approach
is that at large astrophysical and cosmological scales {\it standard general relativity cannot
describe correctly the dynamical evolution of the Universe}. In this context many modified gravity models, all trying to extend and generalized
 GR, have been proposed. Historically, in going beyond the standard gravitational models, the first step was to generalize {\it the geometric part of the Einstein-Hilbert action}. One of the first models of this type is  $f(R)$ gravity, in
which the gravitational action is an arbitrary function of the Ricci scalar $%
R$, so that $S = \frac{1}{16\pi}\int {f(R) \sqrt{-g}d^4 x} +\int {L_m \sqrt{-g}d^4 x}$ \cite{Bu701,Bu702,Bu703,Bu704,Fel}. However, this and many other modifications of the Einstein-Hilbert action concentrate only on the geometric part of the action,
by implicitly assuming that the matter Lagrangian plays {\it a subordinate and passive role} only \cite{Mat}, which naturally follows from its minimal coupling to geometry. But a fundamental theoretical principle forbidding a general coupling between matter and geometry have not been formulated yet, and in fact it does not exist {\it a priori}. If such matter-geometry  couplings are allowed, many theoretical gravitational models with extremely interesting properties can
be constructed.

The first of these kind of models was the $f\left( R,L_{m}\right) $  modified gravity theory \cite{fL1,fL2,fL3,fL4}, with a gravitational action of the form $S = \frac{1}{16\pi}\int {f\left(R,L_m\right) \sqrt{-g}d^4 x} $. A similar geometry-matter type coupling is assumed in the
$f(R,T)$ \cite{fT1,fT2} gravity theory, where $T$ is the trace of the energy-momentum tensor. For a recent review of the generalized $f\left(
R,L_{m}\right)$ and $f(R,T)$ type
gravitational theories with non-minimal curvature-matter coupling see \cite{Revn}. Several other gravitational theories involving geometry couplings have also been proposed, and extensively investigated, like, for example, the
Weyl-Cartan-Weitzenb\"{o}ck (WCW) gravity theory \cite{WCW}, hybrid metric-Palatini $%
f(R,\mathcal{R})$ gravity, where $\mathcal{R}$ is the Ricci scalar
formed from a connection independent of the metric  \cite{HM1,HM2}, $%
f\left(R,T,R_{\mu \nu }T^{\mu \nu }\right)$ type models, where $R_{\mu \nu
} $ is the Ricci tensor, and $T_{\mu \nu }$ the matter energy-momentum
tensor, respectively \cite{Har4,Odin}, or $f(\tilde{T},\mathcal{T})$ gravity \cite{HT}, in which a
coupling between  the torsion scalar $\tilde{T}$ and  the trace of the matter
energy-momentum tensor is assumed.  For a review of hybrid
metric-Palatini gravity see \cite{Revn1}.

 Modified gravity models with geometry-matter coupling are important since they can
provide from a fundamental theoretical point of view a complete theoretical explanations for the late time acceleration of the
Universe, without postulating the existence of dark energy. They can also
offer some alternative explanations for the nature of dark matter. Moreover, these models show that matter itself may play a more fundamental
role in the description of the gravitational dynamics that usually assumed \cite{Mat}, and they can also represent a bridge connecting the classical and the quantum worlds. For example, the dependence of the gravitational action on the trace of the energy-momentum tensor $T$ may be due to the
presence of quantum effects (conformal anomaly), or of some exotic imperfect quantum fluids \cite{Revn}.

Besides the difficulties presented by the present day cosmological observations, a central theoretical problem in present they physics is {\it the unification of quantum mechanics and gravitation}. Gravitation dominates the dynamics of objects at large scales, while quantum mechanics describes the microscopic behaviors of the particles. The study the Universe as a whole from the quantum mechanical point of view is the subject of quantum cosmology \cite{Bo1,bojowald2015quantum}, which is based on {\it the idea that quantum physics must apply to anything in nature, including the whole Universe}. An unification of the electromagnetic force, of the strong force and of the weak force, respectively,  is achieved in the standard model of particle physics, leaving the gravitational force as an exception that cannot be yet unified with the other fundamental forces. This is related to the fact that when considering gravitation in the framework of general relativity,  we must consider not only matter, but also space and time, as physical objects. Space and time obey dynamical laws, and they have excitation such as gravitational waves that interact with each other and with matter. These aspects make quantizing the Universe far from being straightforward. Since the formation of cosmic structures is strongly dependent on the space-time interaction, quantum cosmology is therefore closely related to quantum gravity, representing the quantum theory of the gravitational force and of space-time \cite{Muk}.

 Even being of a speculative and controversial nature, having several difficult conceptual problems to overcome,  quantum cosmology has a long history \cite{DW1,DW2,DW3}, and various popular competing attempts have been proposed to quantized the gravitational field, like, for example, string theory, canonical quantum gravity and loop quantum gravity \cite{Bo1,bojowald2015quantum}. However, the lack of related observations reduces our abilities to resolve conceptual issues to all practical purposes.  Since at the beginning of the Universe the average radius of each point is infinitely small, while the geometric curvature is infinitely large, quantum gravitational effects will dominate the dynamics of the Universe, and therefore they cannot be  neglected  in the study of the very early Universe.  One of the main obstacles in the understanding of quantum cosmology is the so-called {\it problem of time}, which comes from the Hamiltonian constraint in the Arnowitz-Deser-Misner (ADM) formalism, leading to the Wheeler-de Witt equation \cite{DW1,DW2,DW3}, a fundamental equation in canonical quantization of cosmology. The quantum cosmology of $f(R)$ gravity theory  with Schutz's fluid is discussed in \cite{Vak, Vak1}, with new perspectives to the time problem in quantum gravity considered.

 It is the purpose of the present paper to study the quantum cosmology of the $f(R,T)$ gravity theory. Classical aspects of this theory have been extensively investigated \cite{f1,f2,f3,f4,f5,f6,f7,f8,f9,f10,f11,f12,f13}, but its quantum implications have not been considered yet. In order to construct the quantum cosmology of the $f(R,T)$ gravity we adopt for the classical background metric the Friedmann--Robertson--Walker form, and we assume that the matter content of the very early Universe consists of a perfect fluid, described by two thermodynamic parameters only, the energy density, and the thermodynamic pressure, respectively.  In order to introduce the canonical quantization scheme for the $f(R,T)$ gravity theory, as a first step in our study we obtain the general form of the gravitational Hamiltonian, of the quantum potential, and of the canonical momenta, respectively. Once these quantities are explicitly found we write down the full Wheeler-de Witt equation of the $f(R,T)$ modified gravity theory, which describes the quantum properties of the very early Universe, when quantum effects had a dominant influence on the dynamic evolution of the system. We introduce and consider in detail the quantum cosmological properties of a  particular model of the $f(R,T)$ theory, namely, the quantum cosmology of the $f(R,T)=F^0(R)+\theta RT$ model, in which $F^0(R)$ is an arbitrary function of the Ricci scalar, and $\theta $ is a function of the scale factor only. We obtain the Hamiltonian form of the classical equations of motion for this model, and then we write down the  Wheeler-de Witt equation, describing the evolution of the wave function of the early Universe. Starting from the Wheeler-de Witt equation we introduce a time parameter for the corresponding quantum dynamical system, which allows us to formulate
the Schr\"{o}dinger--Wheeler--de Witt equation for the quantum-mechanical model under consideration. In order to obtain the properties of the early Universe we develop  perturbative approach for the study of this cosmological equation, and the energy levels of the Universe are obtained by using a twofold degenerate perturbation approach. Finally, we address the problem of the quantum time by introducing a second quantization approach for the description of the time in the quantum cosmology of the $f(R,T)$ modified gravity theory.

The present paper is organized as follows. The Hamiltonian formulation of the $f(R,T)$ gravity theory is presented in Section~\ref{sect2}, where the canonical momenta associated to the field variables as well as the potential are obtained. This allows us to write down the Wheeler--de Witt equation. The quantum cosmological formulation of the particular model $f(R,T)= F^0(R)+\theta RT$ is presented in Section~\ref{sect3}. In particular, we identify the canonical momentum associated to the time variable, which allows us to transform the Wheeler--de Witt equation into an equivalent Schr\"{o}dinger--Wheeler--de Witt equation. The solutions of this Schr\"{o}dinger--Wheeler--de Witt equation are studied, by using a perturbative approach, in Section~\ref{sect4}. We consider the problem of the second quantization of time in Section~\ref{sect5}. In Section~\ref{sect6} we discuss and conclude our results. The derivation of the field equations of the $f(R,T)$ gravity theory, of the energy balance equation, as well as the canonical  momenta  for the $f(R,T)=F^0(R)+\theta RT$ model, and the expressions of the Ricci tensor for the Friedmann-Robertson-Walker metric are presented in Appendices~\ref{App1}-\ref{App4}, respectively.

 \section{The Wheeler-de Witt equation in $f(R,T)$ gravity theory}\label{sect2}

 We begin our study of the quantum cosmological aspects of the $f(R,T)$ gravity theory by briefly presenting the classical field and conservation equations. Then we obtain the Hamiltonian formulation of the theory, which allows us to write down the Wheeler--de Witt equation in the $f(R,T)$ theory, which describes the quantum evolution of the very early Universe in the presence of geometry-matter coupling. In the present paper we use the natural system of units with  $c = \hbar =16\pi G=1$.

\subsection{The $f(R,T)$ gravity theory}

The gravitational action for the $f(R,T)$ gravity model is \cite{fT1}
\begin{equation}\label{act}
S=\int f(R,T) \sqrt{-g}d^4 x +\int L_m \sqrt{-g}d^4 x.
\end{equation}
We define the energy momentum tensor of the matter as
\begin{equation}
T_{\mu\nu} =-\frac{2}{\sqrt{-g}}\frac{\delta (\sqrt{-g}L_m)}{\delta g^{\mu\nu}}.
\end{equation}

By  assuming that the matter Lagrangian $L_m$ does not depend on the derivative of $g^{\mu\nu}$, we obtain
\begin{equation}
T_{\mu\nu} =g_{\mu\nu} L_m -2\frac{\partial L_m}{\partial g^{\mu\nu}}.
\end{equation}

The field equations of our model can then be obtained as (see Section~\ref{App1} for details)
\bea\label{f1}
\hspace{-0.5cm}&&f_R(R,T) R_{\mu\nu} -\frac{1}{2} f(R,T) g_{\mu\nu} +(g_{\mu\nu}\Box -\nabla_{\mu}\nabla_{\nu}) f_R{R,T}= \nonumber\\
\hspace{-0.5cm}&& \frac{1}{2} T_{\mu\nu} -f_T (R,T) T_{\mu\nu} -f_T (R,T) \Theta _{\mu\nu},
\eea
where $\Theta_{\mu\nu} \equiv g^{\alpha\beta} \delta T_{\alpha\beta}/ \delta g^{\mu\nu}$. Contracting the above field equations we find
\bea\label{f2}
&&f_R (R,T)R -3\Box f_R (R,T) -2f(R,T)=
\frac{1}{2} T -\nonumber\\
&&f_T (R,T) T -f_T (R,T) \Theta .
\eea

For perfect fluid, in the co-moving frame,  the energy-momentum tensor takes the form
\begin{equation}
T_{\mu\nu}=(\epsilon + p)U_{\mu}U_{\nu}+pg_{\mu\nu},
\end{equation}
where $\epsilon $ and $p$ are the matter energy-density and thermodynamic pressure, respectively, and $U^{\mu}$ is the four-velocity, satisfying the normalization condition  $g^{\mu\nu}U_{\mu}U_{\nu} =-1$.

For the perfect fluid we can fix the matter Lagrangian $L_m$ as  $L_m = p$, which gives
\be
\Theta_{\mu\nu} =-2T_{\mu\nu}+pg_{\mu\nu}.
\ee
The $f(R,T)$ gravity theory is a non-conservative theory, and for the energy balance equation we obtain (see Appendix~\ref{App2} for the details of the calculations)
\bea  \label{ther}
&&U_{\mu}\nabla^{\mu}\epsilon +(\epsilon +p)\nabla ^{\mu}U_{\mu} =-\frac{f_T}{\frac{1}{2} +f_T }\times \nonumber\\
&&
 \Bigg[(\epsilon +p)U_{\mu} \nabla^{\mu}\ln f_T +\frac{1}{2}U_{\mu}\nabla^{\mu}(\epsilon -p)\Bigg].\nonumber\\
\eea

\subsection{The effective cosmological Lagrangian  and the potential in $f(R,T)$ gravity theory}

We assume that the geometry of space-time is described by the Friedmann-Robertson-Walker (FRW) metric, which in spherical coordinates is given by
\begin{equation}\label{FRW}
ds^2 = -N^2(t)dt^2 +a^2 (t) \left[\frac{dr^2}{1-kr^2}+r^2 (d\theta^2 +\sin ^2 \theta d\varphi ^2)\right],
\end{equation}
where $N(t)$ is the lapse function, $a(t)$ is the cosmological scale factor, and $k=1,0,-1$,  correspond to the closed, flat and open Universe models, respectively.  To proceed further, as a first step we obtain the effective Lagrangian for the $f(R,T)$ theory,  whose variation with respect to its dynamical variables yields the appropriate equations of motion.

 The trace of the energy-momentum tensor is $T =-\epsilon +3p$. In the comoving reference frame the components of the four-velocity are $U_{\mu}=(N(t),0,0,0)$ and $U^{\mu}=\left(-1/N(t),0,0,0\right)$, respectively. Therefore the trace of the field equation Eq.~(\ref{f2}) of the $f(R,T)$ gravity model becomes
\bea
&&f_R(R,T) R -2f(R,T)+ 3\Box f_R(R,T) =\frac{1}{2} T +\nonumber\\
&&f_T(R,T) T -4p f_T(R,T).
\eea

 With the use of the above identity, and by taking into account the explicit form of the components of the Ricci tensor, given in Appendix~\ref{App3},  we obtain the cosmological action for the $f(R,T)$ gravity theory as
\begin{widetext}
\bea\label{L}
S_{grav}&=&\int dt \Bigg\{Na^3 f(R,T)-\lambda \left[ R-\frac{6}{N}\left(\frac{\ddot{a}}{a}+\left(\frac{\dot{a}}{a}\right)^2 +\frac{kN^2}{a^2}-\frac{\dot{N\dot{a}}}{Na}\right)\right]-\nonumber\\
&&\mu \Bigg[\frac{1}{2} T +f_T(R,T)T-f_R(R,T)R -3\Box f_R(R,T)+2f(R,T)-4pf_T(R,T)\Bigg] \Bigg\}.
\eea
\end{widetext}
 In Eq.~(\ref{L}) $\lambda$ and $\mu$ are {\it Lagrange multipliers}. The term with the second Lagrange multiplier is chosen as the contracted field equation, because it is derived directly from the action, and no further assumptions are needed to be introduced. Moreover, if we use other formulations for the action,  like adopting for the factor multiplying the second Lagrange multiplier the form $T + \epsilon -3p$,  we will lose important insights from the geometry of the modified gravity, and we will have to face conceptual problems when $\epsilon$ and $p$ are related by the radiation equation of state $p=\epsilon /3$.

After taking the variation of Eq.~(\ref{L}) with respect to $R$ and $T$, we obtain the expressions of $\lambda $ and $\mu $ as follows
\bea
\frac{\mu}{Na^3} &=& \frac{f_T}{1/2 +3f_T +f_{TT}T -f_{RT}R -3\Box f_{RT}-4pf_{TT}} \nonumber\\
&\equiv & \widetilde{\mu},
\eea
\bea
\frac{\lambda}{Na^3} &=& f_R -\frac{\mu (f_{RT} T -f_{RR} R + f_R -3\Box f_{RR} -4pf_{RT})}{Na^3} \nonumber\\
&\equiv & \widetilde{\lambda},
\eea

Hence we obtain the gravitational part of the Lagrangian as
\begin{equation}
\mathcal{L}_{grav} = -\frac{6}{N} a \dot{a}^2 \widetilde{\lambda} -\frac{6}{N}a^2 \dot{a}\dot{\widetilde{\lambda}} +6kNa\widetilde{\lambda}- Na^3 V,
\end{equation}
where  the potential $V$ reads
\bea
V& =& -f(R,T) +\widetilde{\lambda} R +\widetilde{\mu} \Bigg[\frac{1}{2} T +f_T(R,T)T-f_R(R,T)R\nonumber\\
&&-3\Box f_R(R,T)+2f(R,T)-4pf_T(R,T)\Bigg].
\eea

In order to make the presentation simpler, we introduce the following notations
\bea
f_R &=& A, f_T =B, f_{RR} = C, f_{RT} = D, f_{TT} = E,\nonumber\\
&& \Box f_{RR} = F, \Box f_{RT} = G,T-4p=M.
\eea
Now $\widetilde{\lambda}$ can be expressed using the new variables as
\begin{equation}
\widetilde{\lambda} = A - \frac{B(DM-CR+A-3F)}{1/2 +3B +EM-DR-3G}.
\end{equation}
In the following we further denote
\begin{equation}
\mathcal{A}=\frac{1}{2} +3B +EM-DR-3G,
\end{equation}
\begin{equation}
\mathcal{Z}=DM-CR+A-3F.
\end{equation}
Thus we have
\begin{equation}
\widetilde{\lambda}= A - \frac{B\mathcal{Z}}{\mathcal{A}},
\end{equation}
After taking the time derivative, we find
\begin{equation}
\dot{\widetilde{\lambda}}=\dot{A} -\frac{B\dot{\mathcal{Z}}+\dot{B}\mathcal{Z}}{\mathcal{A}}+\frac{B\mathcal{Z}\dot{\mathcal{A}}}{\mathcal{A}^2}.
\end{equation}

\subsection{The cosmological Hamiltonian in $f(R,T)$ gravity theory}

The canonical momentum associated to the coordinate $q$ is given by $P_{q}=\frac{\partial\mathcal{L}}{\partial \dot{q}}$. Hence the cosmological Hamiltonian of the $f(R,T)$ gravity is given by
\bea
H_{grav} &&= \dot{a}P_a +\dot{A} P_A + \dot{B}P_B +\dot{C}P_C+\dot{D}P_D+\dot{E}P_E+\nonumber\\
&&\dot{F}P_F+\dot{G}P_G+\dot{R}P_R+\dot{M}P_M+ -\mathcal{L}_{grav}
\eea
After some simple calculations we obtain the explicit forms of the canonical momenta in the cosmological $f(R,T)$, which are presented in Appendix~\ref{App4}.
With these canonical momenta, we obtain the cosmological Hamiltonian of the $f(R,T)$ gravity theory,
\begin{widetext}
\bea
H_{grav} &=& \left(-\frac{6}{N} a \dot{a}^2  \widetilde{\lambda} -6kNa\widetilde{\lambda} + Na^3 V\right) -\frac{6}{N} a^2 \dot{a} \Bigg[\dot{A} -\frac{B\dot{A}}{\mathcal{A}}-\frac{\dot{B}\mathcal{Z}}{\mathcal{A}} +\frac{3B\dot{B}\mathcal{Z}}{\mathcal{A}^2}+\frac{BR\dot{C}}{\mathcal{A}}-\frac{B\dot{D}M}{\mathcal{A}}-\frac{B\mathcal{Z}R\dot{D}}{\mathcal{A}^2}+\nonumber\\
&&\frac{B\mathcal{Z}\dot{E}M}{\mathcal{A}^2}+\frac{3B\dot{F}}{\mathcal{A}}-\frac{3B\mathcal{Z}\dot{G}}{\mathcal{A}^2}+\frac{BC\dot{R}}{\mathcal{A}}-\frac{B\mathcal{Z}D\dot{R}}{\mathcal{A}^2}-\frac{BD\dot{M}}{\mathcal{A}}+\frac{B\mathcal{Z}E\dot{M}}{\mathcal{A}^2} \Bigg].
\eea
\end{widetext}

In order to simplify the notation we will represent the Hamiltonian as $H_{grav}=(\cdots)-\frac{6}{N}a^2 \dot{a}[\cdots]$. Now we can easily find the relation
\bea
P_a P_A &=& \left(\frac{6}{N}\right)^2 a^3 \dot{a}^2 \widetilde{\lambda} \left(1-\frac{B}{\mathcal{A}}\right)+\left(\frac{6}{N}\right)^2 a^4 \dot{a} [\cdots] +\nonumber\\
&&\left(\frac{6}{N}\right)^2 a^4 \dot{a} \left(-\frac{B}{\mathcal{A}}\right)\dot{\widetilde{\lambda}}.
\eea
Since
\begin{equation}
P_a P_F =\left(\frac{6}{N}\right)^2 2a^3 \dot{a}^2 \widetilde{\lambda} \left(3\frac{B}{\mathcal{A}}\right) +\left(\frac{6}{N}\right)^2 a^4 \dot{a} \dot{\widetilde{\lambda}} \left(\frac{3B}{\mathcal{A}}\right),
\end{equation}
by combining the above two equations, we obtain
\begin{equation}
-\frac{6}{N} a^2 \dot{a} [\cdots] = -\frac{N}{6a^2}\left(P_a P_A +\frac{1}{3}P_a P_F\right) + \frac{2\cdot 6}{N} a \dot{a}^2 \widetilde{\lambda}.
\end{equation}
Therefore for the gravitational Hamiltonian we find
\bea
H_{grav}&=& \frac{6}{N}a \dot{a}^2 \left(A-\frac{B\mathcal{Z}}{\mathcal{A}}\right) -\frac{N}{6a^2} \left(P_a P_A +\frac{1}{3}P_a P_F\right)+ \nonumber\\
&& Na^3 V -6kNa \widetilde{\lambda}.
\eea
Since we have $P_A =-\frac{6}{N} a^2 \dot{a}\left(1-\frac{B}{\mathcal{A}}\right)$, after taking the square of it, we have
\bea
\frac{6}{N}a \dot{a}^2 A &=&\left[\frac{P_{A}^{2}}{\left(\frac{6}{N}\right)^2 a^4 \dot{a}^2}-\frac{B^2}{\mathcal{A}^2}+\frac{2B}{\mathcal{A}}\right]\cdot \frac{6}{N} a \dot{a}^2 A =\nonumber\\
&& \frac{NP_A^2}{6a^3} A +\frac{6}{N} a \dot{a}^2 A \left(\frac{-B^2}{\mathcal{A}^2} +\frac{2B}{\mathcal{A}}\right).
\eea

Thus we arrive at the following form of the Hamiltonian,
\bea
H_{grav}&=&\frac{N}{6a^3}P_A^2 A -\frac{N}{6a^2}(P_a P_A+\frac{1}{3}P_a P_F) -\nonumber\\
&&\frac{6}{N}a\dot{a}^2 \frac{AB^2}{\mathcal{A}^2} -\frac{6}{N}a \dot{a}^2 \frac{B(\mathcal{Z}-2A)}{\mathcal{A}}+\nonumber\\
&&Na^3 V -6kNa\widetilde{\lambda}.
\eea
Since
\begin{equation}
P_C P_R = \left(-\frac{6}{N} a^2 \dot{a}\right)^2 \left(\frac{B^2 R C}{\mathcal{A}^2}-\frac{B^2 \mathcal{Z}RD}{\mathcal{A}^3}\right),
\end{equation}
\bea
P_D P_M &=&\left(-\frac{6}{N} a^2 \dot{a}\right)^2 \left(-\frac{BM}{\mathcal{A}}-\frac{B \mathcal{Z}R}{\mathcal{A}^2}\right)\times \nonumber\\ &&\left(-\frac{BD}{\mathcal{A}}+\frac{B\mathcal{Z}E}{\mathcal{A}^2}\right),
\eea
\begin{equation}
\frac{1}{3} P_E P_F =(-\frac{6}{N}a^2\dot{a})^2 \frac{B^2 \mathcal{Z}M}{\mathcal{A}^3},
\end{equation}
\begin{equation}
P_G^2 = \left(-\frac{6}{N}a^2 \dot{a}\right)^2 \frac{9B^2 \mathcal{Z}^2}{\mathcal{A}^4},
\end{equation}
\begin{equation}
P_C \left(P_A +\frac{P_F}{3}\right) =\left(-\frac{6}{N}a^2 \dot{a}\right)^2 \left(\frac{BR}{\mathcal{A}}\right),
\end{equation}
we have
\bea
&&-\frac{6}{N}a \dot{a}^2 \frac{B(\mathcal{Z}-2A)}{\mathcal{A}}= \nonumber\\
&&-\frac{6}{N}a \dot{a}^2 \frac{B\left(DM-CR+A-3F-2A\right)}{1/2 +3B +EM -DR -3G}  = \nonumber\\
&&\frac{N}{6a^3} \left(P_A +\frac{P_F}{3}\right)\frac{P_F}{3} \left( A+3F\right) - \nonumber\\
&&\frac{N}{6a^3} \frac{\mathcal{A}}{B} \left(P_C P_R + P_D P_M\right) +
\frac{6a\dot{a}^2}{N} \frac{\mathcal{A}}{B} \frac{B\mathcal{Z}E}{\mathcal{A}^2}\times \nonumber\\
&&\left(-\frac{BM}{\mathcal{A}}-\frac{B\mathcal{Z}R}{\mathcal{A}^2}\right)-\frac{6a\dot{a}^2}{N}\left(-\frac{2BCR}{\mathcal{A}}\right).
\eea

Therefore we obtain the gravitational part of the cosmological Hamiltonian of the $f(R,T)$ gravity theory as
\bea
H_{grav} &=& \frac{N}{6a^3} \left(P_A +\frac{P_F}{3}\right)(A P_A +F P_F -a P_a) +\nonumber\\
&&\frac{N}{6a^3} \frac{\mathcal{A}}{B} \left(P_C P_R + P_D P_M +\frac{P_E P_F}{3} E+\frac{P_G^2}{9}  RE\right) \nonumber\\
&&+ \frac{N}{6a^3}2P_C \left(P_A +\frac{1}{3}P_F\right)C +Na^3 V-\nonumber\\
&&6kNa \left(A -\frac{B\mathcal{Z}}{\mathcal{A}}\right),
\eea
where the quantum potential $V$ is given by
\begin{equation}
V=-f +\widetilde{\lambda}R +\widetilde{\mu}\left(\frac{1}{2} T +f_T M -f_R R -3\Box f_R +2f \right).
\end{equation}

For the matter part of the Hamiltonian we have \cite{ActionF}
\begin{equation}
H_{matt}=-\mathcal{L}_{matt}= - Na^3 p,
\end{equation}
and thus the total Hamiltonian of the system is
\be\label{H1}
H=H_{grav}+H_{matt}.
\ee

The gravitational Hamiltonian constructed above consists of all canonical momenta associated to all variable of the $f(R,T)$ gravity theory, which can lead to the existence of a complex dynamics of these field variables, and of their associated canonical momenta.

\subsection{The Wheeler-de Witt equation in $f(R,T)$ gravity}

From the Hamiltonian given by Eq.~(\ref{H1}) we immediately obtain the Wheeler-de Witt equation in the framework of the $f(R,T)$ gravity theory  as
\begin{equation}
H \Psi = (H_{grav} +H_{matt})\Psi = N \mathcal{H} \Psi = 0.
\end{equation}

Here the Hamiltonian operator $\mathcal{H}$ takes the form
\bea
\mathcal{H}&=& \frac{1}{6a^3}\Bigg(P_A^2 A + P_A P_F F -P_A P_a a +\frac{A}{3} P_F P_A +\nonumber\\
&&\frac{F}{3}P_F^2 -\frac{a}{3}P_F P_a\Bigg)
- \frac{1}{6a^3} \frac{\mathcal{A}}{B} \Bigg(P_C P_R +P_D P_M  +\nonumber \\
&&\frac{P_E P_F}{3}E +\frac{P_G^2}{9}RE \Bigg) +\frac{2}{6a^3}P_C (P_A +\frac{P_F}{3})C + \nonumber\\
&&a^3 V - 6ka\left(A -\frac{B\mathcal{Z}}{\mathcal{A}}\right) - a^3 p.
\eea

In order to quantize the model we perform first parameter ordering. There are various ways to do it \cite{order}, but in the following we choose a procedure that keeps the Hamiltonian  Hermitian \cite{Vak}. Hence we assume the following relations (we let $P_q = -i (\partial / \partial q)$),
\begin{equation}
q P_{q}^2  = \frac{1}{2}\left(q^u P_q q^v P_q q^w + q^w P_q q^v P_q q^u\right) =  -q\frac{\partial ^2}{\partial q ^2} +u w \frac{1}{q},
\end{equation}
where the parameters $u,v,w$ satisfy $u+v+w=1$, and denote the ambiguity in the ordering of the factors $q$ and $P_q$,
\begin{equation}
q P_q  = \frac{1}{2} \left(q^r P_q q^s + q^s P_q q^r\right) = -i\left(q\frac{\partial}{\partial q}+1\right),
\end{equation}
where the parameters $r,s$ satisfy $r+s =1$, and denote the ambiguity in the ordering of factors $q$ and $P_q$. Similarly we find
\begin{equation}
q^{-2} P_q = -i \left(-\frac{2}{q^3} +\frac{1}{q^2} \frac{\partial}{\partial q}\right).
\end{equation}

Thus we obtain the quantized cosmological Hamiltonian in $f(R,T)$ gravity theory as
\begin{widetext}
\bea
\mathcal{H} &=&-\frac{1}{6a^3}\Bigg[\left(A\frac{\partial}{\partial A} +F\frac{\partial}{\partial F} +2C\frac{\partial}{\partial C} -a\frac{\partial}{\partial a}+5\right)\left(\frac{\partial}{\partial A}+\frac{1}{3}\frac{\partial}{\partial F}\right) -u_1 w_1 \frac{1}{A}-u_2 w_2 \frac{1}{3F}\Bigg]  +\nonumber \\
&&\frac{1}{6a^3}\Bigg[\frac{\mathcal{A}}{B}\left( \frac{\partial}{\partial C}\frac{\partial}{\partial R}  +\frac{\partial}{\partial D}\frac{\partial}{\partial M}+ \frac{E}{3}\frac{\partial}{\partial E} \frac{\partial}{\partial F} - \frac{RE}{9}\frac{\partial ^2}{\partial G ^2} \right) +  \Bigg(- \frac{D}{B}\frac{\partial}{\partial C} -\frac{R}{B}\frac{\partial}{\partial M} +\frac{E}{B}\frac{\partial}{\partial D} +\nonumber\\
&&\frac{\mathcal{A}+EM}{3B}\frac{\partial}{\partial F}   +\frac{RE}{3BG}u_3w_3      \Bigg) \Bigg]+ a^3 V -6ka\left(A -\frac{B\mathcal{Z}}{\mathcal{A}}\right) -a^3 p.
\eea
\end{widetext}

Here $u_1,w_1$, $u_2,w_2$ and $u_3,w_3$ denote the ambiguity in the ordering of factors $A , P_A$, $F, P_F$, and $G,P_G$, respectively.

In the next Section we will investigate some particular quantum cosmological models in the $f(R,T)$ gravity theory.

\section{The quantum cosmology of the  $f(R,T)= F^0(R)+\theta RT$ gravity model}\label{sect3}

After considering the general case of the Wheeler-de Witt equation in the previous Section, we can see that an analytic general solution of this equation for arbitrary $f(R,T)$ would be difficult to obtain. Instead, in the present Section we consider a specific case, in which the gravitational action is of the simple form
\be
f(R,T)= F^0(R) +\theta RT,
\ee
where $F^0(R)$ is an arbitrary function of the Ricci scalar, and $\theta$ is an arbitrary function depending on the scale factor $a(t)$ only. In this toy model, the coupling of the curvature of space-time and the trace of the matter energy-momentum tensor can give a hint of the implications on the quantum cosmological evolution of the existence of such a coupling.

\subsection{The Hamiltonian and the Wheeler-de Witt equation}

In the newly born quantum Universe,  the space-time has a very high curvature, so that $R\rightarrow \infty$. Accordingly, in the present model we have
\be
 \mathcal{A}=\frac{1}{2} +2\theta R, B = \theta R, \frac{B}{\mathcal{A}}=\frac{\theta R}{1/2 +2\theta R}.
 \ee
Similarly, the other variables become
\be
A = F^0_R +\theta T, C = F^0_{RR}, D=\theta , E=0, G=\Box \theta .
\ee

When $R\rightarrow \infty$, $B/\mathcal{A}=1/2$. From the definition of $P_A$ and $P_F$, we obtain $P_F = \left[3B/\left(\mathcal{A}-B\right)\right]P_A \approx 3P_A$. By assuming that in the new born quantum Universe $f_R \gg \Box f_{RR} \rightarrow A \gg  F$, we obtain the gravitational Hamiltonian of the $f(R,T)=F^0(R)+\theta RT$ gravity as
\bea
H_{grav} &=& \frac{N}{6a^3}P_A^2 A \frac{\mathcal{A}}{\mathcal{A}-B}-\frac{N}{6a^2} P_a P_A \frac{\mathcal{A}}{\mathcal{A}-B} -\nonumber\\
&&\frac{N}{6a^3}\frac{\mathcal{A}}{B}\left( P_C P_R +P_D P_M\right) +\frac{2N}{6a^3}\frac{\mathcal{A}}{\mathcal{A}-B} P_A P_C C
 +\nonumber\\
&&Na^3 V -6kNa\left(A-\frac{B\mathcal{Z}}{A}\right)
= -2 \frac{N}{6a^2} P_a P_A +\nonumber\\
&&2\frac{N}{6a^3} P_A^2 A -2\frac{N}{6a^3} \left(P_C P_R +P_D P_M\right)
+ \nonumber\\
&&\frac{4N}{6a^3}P_A P_C C +Na^3 V -6kNa\left(A - \frac{B\mathcal{Z}}{\mathcal{A}}\right).
\eea

In the above equation the quantum potential is defined as
\bea
V&=& \left(\frac{2R\theta}{1/2 +2R\theta}-1\right)\left(F^0-F^0_R R\right)+\nonumber\\
&&\frac{R\theta}{1/2 +2R\theta}\Bigg[\frac{1}{2} T +C R^2 +3F R -3\Box f_R\Bigg].
\eea

In the limit $R\rightarrow \infty$, and by assuming that $\theta \neq 0$, the potential becomes
\begin{equation}
V= \frac{1}{2}\Bigg[\frac{1}{2} T + CR^2 +3F R - 3\Box f_R\Bigg].
\end{equation}

In this approximation the total Hamiltonian of the system becomes
\bea\label{48}
H&=&H_{grav} +H_{matt} = \nonumber\\
&&\Bigg[-\frac{2N}{6a^2}P_a P_A +\frac{2N}{6a^3}P_A^2 A -3kNaA\Bigg]+ \nonumber\\
&&\Bigg[-\frac{2N}{6a^3}P_C P_R +\frac{Na^3}{2}CR^2 -3kNaCR\Bigg]+\nonumber\\
&&\Bigg[-\frac{2N}{6a^3}P_D P_M +\frac{Na^3}{4} M +3kNaDM  \Bigg]+\nonumber\\
&& \frac{4N}{6a^3}P_A P_C C + \frac{3Na^3}{2}\Bigg[F R -\Box f_R \Bigg].
\eea

If in the Lagrangian/ Hamiltonian we have some terms that can be omitted in the action, they can also be omitted in  $L$ and $H$, without causing any physical differences in the dynamics of the system. In Eq.~(\ref{48}, due to Gauss theorem, we have
\bea
\int dt \sqrt{-g} \frac{3Na^3}{2} \Box f_R &=& \int_{M} d^4x \sqrt{-g} \Box f_R \\ &=& \oint _{\partial M}f_R^{;\mu} \sqrt{-g} d\sigma_{\mu}^{3}.
\eea

Then  the variational derivative of this term vanishes,
\begin{equation}
\frac{2}{\sqrt{-g}}\frac{\delta}{\delta g^{\mu\nu}} \int \sqrt{-g}\Box f_R d^4 x = 0.
\end{equation}
Therefore this term can be omitted in the Hamiltonian function. Thus we arrive to the Wheeler-de Witt equation for the $f(R,T)=F^0(R)+\theta RT$ gravitational model, which has the form
\begin{widetext}
\bea
\mathcal{H}\Psi &=& \Bigg\{\left(\frac{2}{6a^2}\frac{\partial ^2}{\partial a \partial A}- \frac{4}{6a^3}C\frac{\partial}{\partial A}\frac{\partial}{\partial C} -\frac{8}{6a^3}\frac{\partial}{\partial A} -\frac{2}{6a^3}A \frac{\partial ^2}{\partial A ^2} +\frac{2}{6}u_1 w_1 \frac{1}{Aa^3}-3kaA\right)+
  \nonumber\\
&&\left(\frac{2}{6a^3}\frac{\partial}{\partial C}\frac{\partial}{\partial R} +\frac{a^3}{2}CR^2 -3kaCR\right)+ \Bigg[\frac{2}{6a^3} \frac{\partial ^2}{\partial D \partial M} +\frac{a^3}{4} M + 3kaDM\Bigg]   +\frac{3a^3}{2}FR\Bigg\}\Psi =0.
\eea
\end{widetext}

\subsection{The Hamiltonian form of the field equations}

In classical mechanics the total time derivative of any function can be obtained with the use of the Poisson bracket $\{,\}$ as
\begin{equation}
\frac{d}{dt}f =\frac{\partial f}{\partial t} + \{f,H \}.
\end{equation}

If the physical variables do not depend explicitly on the time $t$, we obtain
\begin{equation}
\frac{d}{dt}f = \{f,H \}
\end{equation}
Therefore we can formulate the classical equations of motion of the $f(R,T)$ gravity theory as
\begin{equation}
\dot{a} =\{a,H \} = -2\frac{N}{6a^3} P_A,
\end{equation}
\bea
\dot{P_a}&=&\{P_a , H\}=N\Bigg\{\left(-\frac{2}{3a^3}P_A P_a +\frac{1}{a^4}P_A^2 A +3kA\right) +\nonumber\\
&&\left( -\frac{1}{a^4}P_C P_R -\frac{3a^2}{2}CR^2 -3kCR\right) +\Bigg[-\frac{1}{a^4}P_D P_M -\nonumber\\
&&
\frac{3a^2}{4}M +3kDM\Bigg] + \frac{12}{6a^4}P_A P_C C  \Bigg\},
\eea
\begin{equation}
\dot{A}= \{A,H \}=N\Bigg[-\frac{1}{3a^2}P_a +\frac{2}{3a^3}AP_A  +\frac{4}{6a^3}P_C C \Bigg],
\end{equation}
\begin{equation}
\dot{P_A}=\{P_A,H \}=N\Bigg[-\frac{1}{3a^3}P_A^2 +3ka\Bigg],
\end{equation}
\begin{equation}
\dot{C} = \{C,H \}=-\frac{N}{3a^3}\Bigg[(P_R -2P_A C \Bigg],
\end{equation}
\begin{equation}
\dot{P_C} =\{P_C , H \}= N\Bigg[-\frac{a^3}{2}R^2 +3kaR -\frac{4}{6a^3}P_A P_C \Bigg],
\end{equation}
\begin{equation}
\dot{R} = \{R,H \} =-\frac{N}{3a^3}P_C,
\end{equation}
\begin{equation}
\dot{P_R} =\{P_R , H\} = N\Bigg[-a^3 CR + 3kaC -\frac{3a^3}{2}F\Bigg],
\end{equation}
\begin{equation}
\dot{D} =\{D,H \}=-2 \frac{N}{6a^3} P_M,
\end{equation}
\begin{equation}
\dot{P_D} =-3kNaM,
\end{equation}
\begin{equation}
\dot{M} = -2\frac{N}{6a^3} P_D,
\end{equation}
\begin{equation}
\dot{P_M} = -N\left(\frac{a^3}{4} +3kaD\right).
\end{equation}

Now we define a new time variable $\tau$, which has the following relation with the original time variable $t$,
\begin{equation}
\tau = \int N(t) dt,
\ee
or, equivalently, $d\tau/dt = N(t)$. From the definition of $P_M$, we have
\begin{equation}
P_M = \frac{3a^2 \dot{a}}{N}D  = 3a^3 h D.
\end{equation}

where $h \equiv \frac{1}{N} \frac{\dot{a}}{a}$  is the Hubble function. Then the Hamilton equations of motion for a cosmological fluid become
\begin{equation} \label{Dprime}
D^{'} = -h \theta =-\frac{a'}{a} D,
P_D^{'} =-3kaM,
\end{equation}
\begin{equation}
M' = -2 \frac{1}{6a^3} P_D,
P_M^{'} = (\frac{a^3}{4} +3kaD).
\end{equation}
where a prime represents the derivative with respect to  $\tau$. The first equation of the above system  gives us the coupling constant $\theta =D$ as
\begin{equation}
D= \frac{\delta}{a(\tau)} , \qquad \delta = {\rm constant}.
\end{equation}

This result tells us that the coupling between the gravitational field and the matter field decreases as the scale of the Universe increases. This result may be the reason why the coupling between matter and gravity becomes so weak in the limit of large cosmological times, and the $f(R,T)$ gravity behaves as the standard gravity nowadays.

\subsection{ The time problem in the quantum cosmology of the $f(R,T)$ gravity }

The absence of the time evolution of the wave function of the Universe in the Wheeler-de Witt equation hinders our efforts of understanding quantum gravity in a way similar to standard quantum mechanics, or quantum field theory. Below we propose a way to turn the Wheeler-de Witt equation into a Schr\"{o}dinger type equation.  For the product $P_D P_T$ we obtain immediately the result
 \begin{equation}
 -\frac{2}{6a^3} P_D P_M = -h D P_D.
 \end{equation}
 In the following we make the fundamental assumption that the above term can be interpreted as  $P_{\tau}$, which is {\it the canonical momentum for time}. There is a convincing reason for this assumption: after performing quantization $P_q = -i \frac{\partial}{\partial q}$, we obtain
\bea
P_{\tau} &=& -i\frac{d}{d\tau} = \frac{d a(\tau)}{d\tau} \left(-i\frac{\partial}{\partial a}\right)= \\
&& -\frac{a'}{a} \frac{\delta}{a}\left(-\frac{a^2}{\delta}\right)\left(-i \frac{\partial}{\partial a}\right) = -h D P_D.
\eea
This result holds if we assume that $a(t)$ is the only time dependent variable in the model. The relation proved above allows us to perform the transformation
\begin{equation} \label{timeT}
-\frac{2}{6a^3}P_D P_M \rightarrow P_{\tau}.
\end{equation}

The validity of this transformation shows us that the coupling between the gravitational field and the matter field may play an important role in the evolution of the very early Universe. In the following we will make a further simplification of the gravitational action, by assuming it to be
\be
f(R,T) = R+\theta R T.
\ee
Then $C=F= P_R=0$, and for the cosmological Hamiltonian of the system we obtain
 \bea
H&=&H_{grav} +H_{matt} = -\frac{2N}{6a^2}P_a P_A +\frac{2N}{6a^3}P_A^2 A -\nonumber\\
&&3kNaA -\frac{N}{6a^3}P_D P_M +\frac{Na^3}{4} M +3kNaM.
\eea

This Hamiltonian is very similar to the Hamiltonian obtained in the $f(R)$ model presented in \cite{Vak}, except for the appearance of a new term $-2\frac{N}{6a^3}P_D P_T +N(\frac{a^3}{4} +3ka)M $, which shows us the effect of the coupling between space-time and matter.
The Wheeler-de Witt equation $\mathcal{H}\Psi =0$ for this $f(R,T)$ gravity  model reads
\bea
\mathcal{H}\Psi &=& \Bigg[ -\frac{2}{6a^2}P_a P_A +\frac{2}{6a^3}P_A^2 A -3kaA
-\frac{2}{6a^3}P_D P_M +\nonumber\\
&&\frac{a^3}{4}M +3kaDM \Bigg]\Psi =0.
\eea

After the use of the transformation introduced in Eq.~(\ref{timeT}), we obtain
\bea
\mathcal{H}_{eff}\Psi &=& \Bigg[ -\frac{2}{6a^2}P_a P_A +\frac{2}{6a^3}P_A^2 A -3kaA
 +\nonumber\\
 &&\frac{a^3}{4} M +
 3kaDM \Bigg]\Psi = -P_{\tau}\Psi.\nonumber\\
\eea

After substituting all $P_q = -i\frac{\partial}{\partial q}$  to quantize the model, we obtain the corresponding Schr\"{o}dinger-Wheeler-de Witt (SWDW) equation describing the quantum evolution of the Universe as
\bea \label{H}
\mathcal{H}_{eff}\Psi= \Bigg[\Bigg(\frac{2}{6a^2}\frac{\partial ^2}{\partial a \partial A} -\frac{4}{6a^3}\frac{\partial}{\partial A} -\frac{2}{6a^3}A \frac{\partial ^2}{\partial A ^2} +\nonumber \\
\frac{2}{6}u_1 w_1 \frac{1}{Aa^3} -3kaA\Bigg)  +\frac{a^3}{4} M +3kaDM\Bigg]\Psi = i\frac{\partial \Psi}{\partial \tau},
\eea
which is just of the form of the standard Schr\"{o}dinger equation,
\begin{equation}
\mathcal{H}_{eff}\Psi = i\frac{\partial \Psi}{\partial \tau}.
\end{equation}

Therefore in the $f(R,T)$ gravity theory, {\it we can generate a Schr\"{o}dinger type equation from the Wheeler-de Witt equation}, which can solve the time problem in  quantum gravity. When $\frac{d\tau}{dt}= N(t) = 1$, the  WDW equation will take the form of the Schr\"{o}dinger equation we are familiar with,
\begin{equation}
\mathcal{H}\Psi = i\frac{\partial\Psi}{\partial t}.
\end{equation}

Let us take now a deeper look into the time problem of quantum gravity, and analyze the physical meaning of the time $\tau$, and of the effective Hamiltonian $H_{eff}$ we have introduced here. In the Wheeler-de Witt equation $\mathcal{H}\psi =0$ there seems to be no dynamics of the system. Therefore it turns out that the wave function of the Universe (more precisely, the physical states ) do not describe states of quantum gravity at a particular time, as in the standard quantum theory. Rather, they describe states for all times, or, more precisely, just that information about the state of the Universe that is invariant under all space-time diffeomorphisms \cite{GKG}

In the modified gravity model $f(R,T) =R +\theta RT$, time may be introduced locally by the coupling of the gravitational field and the matter field, because the interaction between the gravitational field and the matter field is also local.  The profound connections between thermodynamics and gravity tell us that the arrow of time may come from the second law of thermodynamics, since both processes reveal irreversible dynamics. If we only consider a small pitch of the Universe, and think of it as an adiabatic system, with $\mathcal{H} \Psi =0$ still valid in it, the coupling between curvature and matter will generate an arrow of time to measure the increase of its entropy, contributed by matter. The other components of the Hamiltonian, included in $\mathcal{H}_{eff}$,  show us the dynamics of the gravitational field, and of some matter components, so that the effective Hamiltonian  takes the form $\mathcal{H} _{eff} \sim b_1 p_i p_j + b_2 x_i x_j$, which is similar to the Hamiltonian we usually meet in quantum mechanics. Therefore one can suppose that the Schr\"{o}dinger equation in ordinary quantum mechanics might describe just a locally effective theory of the Wheeler-de Witt equation. In other words, we may conjecture that the Wheeler-de Witt equation provides the global quantum description for the Universe, while the Schr\"{o}dinger equation is just the local description for the present day microscopic regions of the Universe.

\section{ A perturbative approach to the cosmological SWDW equation in $f(R,T)$ gravity}\label{sect4}

In order to solve the SWDW equation Eq.~(\ref{H}) for the wave function of the Universe, we look for stationary solutions, and we separate the variables  as \begin{equation}
\Psi(a,A,D,M,\tau) =e^{-iE\tau} \psi (a,A,D,M)
\end{equation}
Here $E={\rm constant}.$ Thus we obtain the following differential equation describing the time evolution of the wave function of the Universe,
\bea
\mathcal{H}\psi&=& \Bigg[\Bigg(\frac{2}{6a^2}\frac{\partial ^2}{\partial a \partial A} -\frac{4}{6a^3}\frac{\partial}{\partial A} -\frac{2}{6a^3}A \frac{\partial ^2}{\partial A ^2} +\nonumber\\
&&\frac{2}{6}u_1 w_1 \frac{1}{Aa^3} -3kaA\Bigg)  +\frac{a^3}{4}M +\nonumber\\
&&3kaDM -E \Bigg]\psi = 0.
\eea
By introducing the new variables $x=aA^{\frac{1}{2}}$ and $y = A$, we obtain the differential equation for $\psi$
\bea\label{96}
&&\Bigg[\frac{1}{4}x^2 \frac{\partial^2}{\partial x^2}-\frac{1}{4}x\frac{\partial}{\partial x}-2y\frac{\partial}{\partial y}-y^2 \frac{\partial^2}{\partial y^2} +u_1 w_1 -9kx^4 +\nonumber\\
&&3M\left(\frac{a^6}{4} +3ka^4D\right)y -3x^3 y^{-\frac{1}{2}} E\Bigg]\psi =0.
\eea
In the following we approximate the equation of state of the early Universe by the stiff equation of state $p=\epsilon$, since in the very high density Universe one expects the speed of sound $c_s$ to be of the same order of magnitude as the speed of light, $c_s =\sqrt{\partial \epsilon /\partial p} =1$. When $p=\epsilon$, and since $\dot{R}= -\frac{N}{3a^3}P_C = 2\frac{\dot{a}}{a} \frac{1}{2} R =\frac{\dot{a}}{a} R$, from Eq.~(\ref{ther}) it follows that in this case the energy is conserved, $\epsilon' +3(\epsilon + p) h =0$. Hence we have
\begin{equation}
M= T-4p = -\frac{w}{a^6},
\end{equation}
where $w>0$ is a positive constant. By substituting this result into Eq.~(\ref{96}), and by letting $D =\delta / a(\tau)$, we find
\bea
&&\Bigg[\frac{1}{4}x^2 \frac{\partial^2}{\partial x^2}-\frac{1}{4}x\frac{\partial}{\partial x}-2y\frac{\partial}{\partial y}-y^2 \frac{\partial^2}{\partial y^2} +u_1 w_1 -9kx^4 -\nonumber\\
&&\frac{3w}{4} y -9w\delta k \frac{y^{\frac{5}{2}}}{x^3} -3x^3 y^{-\frac{1}{2}} E\Bigg]\psi =0.
\eea

\subsection{Time evolution as a non-constant energy perturbation}

In order to obtain a solution of  the SWDW equation obtained above, we can estimate different terms' order of magnitude. We know that $x=aA^{1/2} \sim 1/a^{\frac{5}{2}}$, $y = A \sim 1/a^7$, and hence we can obtain
 \begin{equation}
x^4 \sim 1/a^{10} , \qquad y^{\frac{5}{2}}/x^3 \sim 1/a^{8}, \qquad x^3/y^{\frac{1}{2}} \sim 1/a^4.
 \end{equation}
Since $a$ is very small, the term $x^3/y^{\frac{1}{2}} \sim 1/a^4$ can be thought as a small perturbation. To obtain some analytic solutions, we consider the case when $k=0$ (flat Universe).  This choice can help us to get rid of the large coupling term $y^{\frac{5}{2}}/x^3 \sim 1/a^{8}$. Due to the discussion by Hawking and  Page \cite{HP}, we also assume that $u_1 w_1$, the ordering parameter, can be neglected. In our case this can be achieved by putting it into the energy term, or neglecting it directly, since it is small as compared to the variables related to $a(\tau)$.

After making the above assumptions, we obtain the Schr\"{o}dinger-Wheeler-de Witt equation as
\begin{equation}
\frac{y^{\frac{1}{2}}}{3x^3}\Bigg[\frac{1}{4}x^2 \frac{\partial ^2}{\partial x^2} -\frac{1}{4}x\frac{\partial}{\partial x}-2y \frac{\partial}{\partial y}-y^2 \frac{\partial^2}{\partial y^2} -\frac{3w}{4} y\Bigg]\Psi = i\frac{\partial}{\partial \tau} \Psi.
\end{equation}
where $\Psi = \Psi (x,y,\tau)$. In the stationary situation, we can decompose the variables as $\Psi(x,y,\tau) =e^{-iE\tau} \psi (x,y)$, and thus
we get
\begin{equation}\label{ffS}
\frac{y^{\frac{1}{2}}}{3x^3}\Bigg[\frac{1}{4}x^2 \frac{\partial ^2}{\partial x^2} -\frac{1}{4}x\frac{\partial}{\partial x}-2y \frac{\partial}{\partial y}-y^2 \frac{\partial^2}{\partial y^2} -\frac{3w}{4} y\Bigg]\psi = E\psi.
\end{equation}

Now we change the above equation into another form, and multiply by $\frac{3x^3}{y^{\frac{1}{2}}}$  both sides. We thus have the equation
\begin{equation}
\Bigg[\frac{1}{4}x^2 \frac{\partial ^2}{\partial x^2} -\frac{1}{4}x\frac{\partial}{\partial x}-2y \frac{\partial}{\partial y}-y^2 \frac{\partial^2}{\partial y^2} -\frac{3w}{4} y -\frac{3Ex^3}{y^{\frac{1}{2}}}\Bigg]\psi = 0.
\end{equation}

In the first formulation of the SWDW equation, given by Eq.~(\ref{ffS}), we can think of the time evolution of the wave function as resulting in the addition of a constant $E$ as a perturbation. But what we know from the perturbation theory of quantum mechanics tells us that a constant perturbation gives us no changes in the energy level and in the wave function, which is to say that it does not affect the physical observables. After changing the unperturbed Hamiltonian in the following way
\begin{eqnarray*}
&&\frac{y^{\frac{1}{2}}}{3x^3}\left[\frac{1}{4}x^2 \frac{\partial ^2}{\partial x^2} -\frac{1}{4}x\frac{\partial}{\partial x}-2y \frac{\partial}{\partial y}-y^2 \frac{\partial^2}{\partial y^2} -\frac{3w}{4} y \right] \rightarrow\nonumber\\
&&\left[\frac{1}{4}x^2 \frac{\partial ^2}{\partial x^2} -\frac{1}{4}x\frac{\partial}{\partial x}-2y \frac{\partial}{\partial y}-y^2 \frac{\partial^2}{\partial y^2} -\frac{3w}{4} y\right],
\end{eqnarray*}
we obtain the new version of the SWDW equation, where we do not consider the constant $E$ as induced by a time evolution effect anymore. Instead, we consider it as a non-constant perturbation
\be
V_{pert} = -\frac{3Ex^3}{y^{\frac{1}{2}}},
 \ee
of the system, such that the total Hamiltonian is $H = H_0 +V_{pert}$. Hence  we can say that the time evolution effect on the wave function in a non-perturbative system, like the $f(R,T)$ gravity theory, is equivalent to the splitting of the degenerate energy levels (loss of symmetries) in a perturbative, static system. This result shows that the time evolution in the quantum cosmology of $f(R,T)$ gravity leads to the splitting of the degenerate energy levels, which reveals the deep connections between energy and time.

In the new form of the Hamiltonian the unperturbed component $H_0$ satisfies the equation $H_0 \psi =0$, whose eigenvalue of energy is zero. In the unperturbed Hamiltonian we can separate the variables as $\psi(x,y) = X(x)Y(y)$, and obtain
\begin{eqnarray}
\hspace{-0.5cm}&&\Bigg[x^2\frac{\partial ^2}{\partial x^2} -x\frac{\partial}{\partial x} +(1-v^2) \Bigg] X(x) =0, \\
\hspace{-0.5cm}&&\Bigg[y^2 \frac{\partial^2}{\partial y^2} +2y \frac{\partial}{\partial y} +\frac{3w}{4} y -\frac{v^2 -1}{4}\Bigg]Y(y) =0,
\end{eqnarray}
where the separation constant is denoted as $\frac{v^2 -1}{4}$. Then we obtain the expression of the wave function for the unperturbed Hamiltonian $H_0$ as
\bea
\Psi (x,y,\tau) &=&e^{-iE\tau} \left(A_1 x^{1-v} + A_2 x^{-1-v} \right) \frac{1}{\sqrt{y}}\times \nonumber\\
&&\left(B_1 J_v (\sqrt{3w y}) +B_2 J_{-v} (\sqrt{3w y})\right),\nonumber\\
\eea
where $A_1$, $A_2$, $B_1$, $B_2 $ are integration constants. $J_v(x)$ is the Bessel function, and $J_v (x)$ and $J_{-v}(x)$ are linearly independent functions. Since in the early Universe, we have $x\rightarrow \infty$, in order to have analytic solution in the whole plane, we let $A_1= B_2 =0$, and we assume $v \ge 0$.

\subsection{The twofold degenerate case}

 In order to investigate the energy level split, we consider the simplest case, the twofold degenerate case \cite{LDQM}. Assuming that the system is twofold degenerate,  at the beginning of time $\tau = 0$, the wave function can be written as
 \begin{equation}
 \Psi = c_1 \psi_1 + c_2 \psi_2
 \end{equation}
where $c_1, c_2$ are constants satisfying the relation $\vert c_1 \vert ^2 +\vert c_2 \vert ^2 =1$, and
\begin{eqnarray}
\psi_1 &=&x^{-1-v_1}\frac{1}{\sqrt{y}} J_{v_1} \left(\sqrt{3w y}  \right),  \\
\psi_2 &=&x^{-1-v_2}\frac{1}{\sqrt{y}} J_{v_2} \left(\sqrt{3w y}  \right),
\end{eqnarray}
where $v_1$ and $v_2$ are positive constants. For the  perturbed system with perturbation  $V_{pert} =-3E\frac{x^3}{y^{\frac{1}{2}}}$, we write
\begin{equation}
V_{ij} =\int \psi_i^{*} V_{pert} \psi _j dx dy.
\end{equation}
With the use of the unperturbed wave functions, we obtain
\bea
V_{11}&=&\int \psi_1^{*} V_{pert} \psi_1 dxdy =\nonumber\\
&&\int (-3E)x^{1-2v_1} \frac{1}{y^{\frac{3}{2}}}[J_{v_1} (\sqrt{3w y})]^2 dx dy ,
\eea
\bea
V_{22}&=&\int \psi_2^{*} V_{pert} \psi_2 dxdy =\nonumber\\
&&\int (-3E)x^{1-2v_2} \frac{1}{y^{\frac{3}{2}}}[J_{v_2} (\sqrt{3w y})]^2 dx dy,
\eea
\bea
V_{12}&=&V_{21}^{*}=\int \psi_1^{*} V_{pert} \psi_2 dxdy =\int (-3E)x^{1-v_1 -v_2} \times \nonumber\\
&&\frac{1}{y^{\frac{3}{2}}}[J_{v_1} (\sqrt{3w y})J_{v_2} (\sqrt{3w y})] dx dy.
\eea

Since the curvature scalar and the scale factor are always positive, we cannot define wave functions that are analytic on whole space, thus the orthogonality and normalization are not satisfied. In the following we define the quantities
\begin{equation}
S_{11} = \int \psi^{*}_1 \psi_1 = \int x^{-2-2v_1}\frac{1}{y} [J_{v_1}(\sqrt{3w y}) ]^2 dxdy,
\ee
\be
S_{22} = \int \psi^{*}_2 \psi_2 = \int x^{-2-2v_2}\frac{1}{y} [J_{v_2}(\sqrt{3w y}) ]^2 dxdy,
\ee
\bea
S_{12}& =&S_{21}^{*} = \int \psi^{*}_1 \psi_2 =\nonumber\\
&&\int x^{-2-v_1-v_2}\frac{1}{y} J_{v_1}(\sqrt{3w y}) J_{v_2}(\sqrt{3w y})dxdy.\nonumber\\
\eea

In the special function theory, we have the Schafheitlin integral \cite{SF}, which reads
\bea
&&\int ^{\infty}_{0} \frac{J_{\mu}(at) J_{\nu}(bt)}{t^{\lambda}} dt =\nonumber\\
&&\frac{\Gamma(\lambda)\Gamma(\frac{\mu +\nu -\lambda +1}{2})(\frac{a}{2})^{\lambda -1}}{2\Gamma(\frac{\mu-\nu+\lambda +1}{2}) \Gamma(\frac{\nu -\mu +\lambda+1}{2}) \Gamma(\frac{\mu+\nu+\lambda +1}{2})},
\eea
where $a$ and $b$ are positive constants, and satisfy the relations below to make the integral convergent,
\begin{equation}
{\rm Re} (\mu +\nu +1) > {\rm Re}(\lambda) >0.
\end{equation}

For $z \in \Re$ the Gamma function $\Gamma (z)$ satisfies  the identities $\Gamma (z+1) =z\Gamma (z) , \Gamma(1)=1, \Gamma(\frac{1}{2})=\sqrt{\pi}$. In the following we will restrict our analysis to the simplest case $v_1 = \frac{3}{2}, v_2 =\frac{5}{2}$. With the use of these values we obtain
\bea
V_{11}&=& -3E \int \frac{1}{x^2} \frac{[J_{\frac{3}{2}} (\sqrt{3w y})]^2}{y^{\frac{3}{2}}}dxdy =\nonumber\\
&&\eta \int^{\infty}_{l}dx \int^{\infty}_{0} \frac{1}{x^2} \frac{[J_{\frac{3}{2}} (z)]^2}{z^2}dz
= \nonumber\\
&&\eta \int^{\infty}_{l} \frac{dx}{x^2} \frac{1}{2\pi} = \frac{\eta}{2\pi l},
\eea
\bea
V_{22}&=& -3E \int \frac{1}{x^4} \frac{[J_{\frac{5}{2}} (\sqrt{3w y})]^2}{y^{\frac{3}{2}}}dxdy =\nonumber\\
&&\eta  \int^{\infty}_{l}dx \int^{\infty}_{0} \frac{1}{x^4} \frac{[J_{\frac{5}{2}} (z)]^2}{z^2}dz
= \nonumber\\
&&\eta \int^{\infty}_{l} \frac{dx}{x^4} \frac{1}{6\pi} = \frac{\eta}{18\pi l^3},
\eea
\bea
V_{12}&&= V_{21} = -3E \int \frac{1}{x^3} \frac{J_{\frac{3}{2}} (\sqrt{3w y}) J_{\frac{5}{2}} (\sqrt{3w y})}{y^{\frac{3}{2}}}dxdy =\nonumber\\
&&\eta  \int^{\infty}_{l}dx \int^{\infty}_{0} \frac{1}{x^3} \frac{J_{\frac{3}{2}} (z) J_{\frac{5}{2}} (z)}{z^2}dz
= \nonumber\\
&&\eta \int^{\infty}_{l} \frac{dx}{x^3} \frac{1}{15} = \frac{\eta}{30 l^2},
\eea
\bea
S_{11} &= &\int \frac{1}{x^5} \frac{[J_{\frac{3}{2}}(\sqrt{3w y})]^2}{y} dxdy  = \nonumber\\
&& \int^{\infty}_{l}dx \int^{\infty}_{0} \frac{2}{x^5} \frac{[J_{\frac{3}{2}} (z)]^2}{z}dz
= \nonumber\\
&&\int^{\infty}_{l} \frac{2dx}{x^5} \frac{1}{3} = \frac{1}{6 l^4},
\eea
\bea
S_{22} &=& \int \frac{1}{x^7} \frac{[J_{\frac{5}{2}}(\sqrt{3w y})]^2}{y} dxdy \nonumber \\&=&  \int^{\infty}_{l}dx \int^{\infty}_{0} \frac{2}{x^7} \frac{[J_{\frac{5}{2}} (z)]^2}{z}dz =\nonumber\\
&& \int^{\infty}_{l} \frac{2dx}{x^7} \frac{1}{5} = \frac{1}{15 l^6} ,
\eea
\bea
S_{12} &&= S_{21} = \int \frac{1}{x^6} \frac{J_{\frac{3}{2}}(\sqrt{3w y})J_{\frac{5}{2}}(\sqrt{3w y})}{y} dxdy = \nonumber\\
&& \int^{\infty}_{l}dx \int^{\infty}_{0} \frac{2}{x^6} \frac{J_{\frac{3}{2}} (z) J_{\frac{5}{2}} (z)}{z}dz
= \nonumber\\
&&\int^{\infty}_{l} \frac{2dx}{x^6} \frac{1}{2\pi} = \frac{1}{5\pi l^5},
\eea
where we have denoted $z= \sqrt{3w y} \rightarrow dz =\frac{\sqrt{3w}}{2\sqrt{y}}dy$, and  $\eta = -6E \sqrt{3w}$, respectively, and we have also assumed that $x$ and $y$ are independent variables. The upper limits in the integrals are both $\infty$ when $a(\tau) \rightarrow 0$. As for the lower limits of integration we assumed them to be very small positive numbers $l$ and $l_1$, which correspond to the transition from the quantum regime to the classical regime, and thus they describe the limit of applicability of the present quantum model of the Universe. From the numerical evaluation of the integrals it follows that when $l_1 <0.5$ the integral values will not vary much, and therefore we let $l_1 =0$.

\subsection{The energy levels of the quantum Universe}

We use the degenerate perturbation theory to find the energy levels in the quantum cosmology of $f(R,T)$ gravity.  The wave function is given by  $\Psi =c_1 \psi +c_2 \psi_2$, and we already know the eigenvalues and the eigenfunctions of the unperturbed Hamiltonian $H_0$, which are given as solutions of the equation
\begin{equation}
H_0 \psi_i =E^{(0)} \psi_i, \qquad i=1,2.
\end{equation}

By substituting the above equation in the Schr\"{o}dinger equation, $H\Psi = (H_0 +V) \Psi =E\Psi$, we obtain
\begin{equation}\label{Sch1}
(H_0 +V)(c_1 \psi_1 +c_2 \psi_2) =E(c_1 \psi_1 +c_2 \psi_2).
\end{equation}
 After  multiplying Eq.~(\ref{Sch1}) with $\psi_1^{*}$ and $\psi_2^{*}$, and integrating over a volume $V$, we obtain the equations
\bea
&&c_1 \left(E^{(0)}_1 S_{11} +V_{11} -ES_{11}\right) +\nonumber\\
&&c_2\left(E^{(0)}_2 S_{12} +V_{12} -ES_{12}\right)= 0,
\eea
\bea
&&c_1 \left(E^{(0)}_1 S_{21} +V_{21} -ES_{21}\right) +\nonumber\\
&&c_2\left(E^{(0)}_2 S_{22} +V_{22} -ES_{22}\right)= 0.
\eea
In the degenerate situation, we let $E=E^{(0)} +E^{(1)}$, and $c_n = c_n^{(0)}$, that is, we take for these coefficients the zero order (unperturbed) approximation.  Then we have
\begin{equation} \label{c1c2}
c_1 \left(V_{11} -E^{(1)} S_{11}\right) +c_2\left(V_{12} -E^{(1)} S_{12}\right)= 0,
\ee
\be
c_1 \left(V_{21} -E^{(1)} S_{21}\right) +c_2\left(V_{22} -E^{(1)} S_{22}\right)= 0.
\end{equation}

From these two equations we can find the coefficients $c_1,c_2$, after the perturbed energy $E^{(1)}$ is obtained. The secular equation is
\begin{equation}
\left\vert
\begin{array}{cc}
V_{11} -E^{(1)} S_{11} & V_{12} -E^{(1)} S_{12}\\
V_{21} -E^{(1)} S_{21} & V_{22} -E^{(1)} S_{22} \\
\end{array}
\right\vert
=0.
\end{equation}

Then we obtain the first order modifications of the energy as
\begin{equation}
E^{(1)} = \frac{-B \pm \sqrt{B^2 -4AC}}{2A},
\end{equation}
where
\begin{equation}
A = S_{11}S_{22} -S_{12}S_{21},
\ee
\be
B = S_{12}V_{21} + S_{21}V_{12} -S_{11}V_{22} -S_{22}V_{11},
\ee
\be
C = V_{11}V_{22} -V_{12}V_{21}.
\end{equation}

In the case analyzed earlier, where $v_1 = \frac{3}{2}, v_2 = \frac{5}{2}$, we obtain
\begin{equation}
A = \left(\frac{1}{90}-\frac{1}{25\pi^2}\right)\frac{1}{l^{10}},
B = -\frac{79}{2700\pi} \frac{\eta}{l^7},
\ee
\be
C = \left(\frac{1}{36\pi^2} -\frac{1}{900}\right)\frac{\eta^2}{l^4}.
\end{equation}

And the two modified energy levels are
\bea
&&E^{(1)}_{\pm} =\frac{-B \pm \sqrt{B^2 -4AC}}{2A} =\nonumber\\
 &&\frac{ \frac{79}{2700\pi} \pm \sqrt{\left(\frac{79}{2700\pi}\right)^2 -4\left(\frac{1}{90} -\frac{1}{25\pi^2}\right)\left(\frac{1}{36\pi^2} -\frac{1}{900}\right)}}{2\left(\frac{1}{90}-\frac{1}{25\pi^2}\right)}\eta l^3  \nonumber\\
&&= \beta_{\pm}\eta l^3 \approx ( 0.660\pm 0.440) \eta l^3
\eea
where $\beta_{\pm}$ are two constants. From the above equation we know that the modified energies have the same sign, and they are proportional to $E$,  as well as proportional to $l^3$, which will tell us the lower limit of the size of the energy gap, if we know the numerical value of  $l$.
From Eq~(\ref{c1c2}) the wave function's coefficients are obtained as
\bea
c_1 &=& \sqrt{\frac{\left(\frac{\eta}{30 l^2} -E^{(1)}_{\pm}\frac{1}{5\pi l^5}\right)^2}{\left(\frac{\eta}{2\pi l} -E^{(1)}_{\pm}\frac{1}{6l^4}\right)^2 +\left(\frac{\eta}{30 l^2} -E^{(1)}_{\pm}\frac{1}{5\pi l^5}\right)^2} }= \nonumber\\
&&\sqrt{\frac{\left(\frac{1}{30} -\frac{\beta_{(\pm)}}{5\pi}\right)^2}{\left(\frac{1}{2\pi} -\frac{\beta_{\pm}}{6}\right)^2 l^2 + \left(\frac{1}{30} -\frac{\beta_{\pm}}{5\pi}\right)^2} },
\eea{equation}
and
\begin{equation}
c_2 = \pm \sqrt{\frac{\left(\frac{1}{2\pi} -\frac{\beta_{\pm}}{6}\right)^2 l^2}{\left(\frac{1}{2\pi} -\frac{\beta_{\pm}}{6}\right)^2 l^2 + \left(\frac{1}{30} -\frac{\beta_{\pm}}{5\pi}\right)^2} },
\end{equation}
respectively.

The above expressions show the dependence of the coefficients $c_1$ and $c_2$ on $l$. When $l$ is large, the Universe will have a higher probability to be in the state $\psi_2$, and it will have a higher probability to be in the state $\psi_1$ when $l$ is small. Note that the coefficients $c-1$ and $c_2$ do not depend on $\eta$, which implies that the parameter $E$ will not affect the state of the wave function.

\subsection{The transition probability in the quantum Universe}

 In the previous Section we have considered the early Universe as a quantum  system that has twofold degenerate energy levels corresponding to the wave functions  $\psi_1, \psi_2$ at the time $\tau = 0$. In the following we will consider the probability of transition of the Universe, from the $\psi_1$ state at $\tau =0$, to the $\psi_2$ state at time $\tau$, the transition taking place due to a perturbation of the initial state.

In the zero order approximation the wave functions are
\begin{equation}		
\psi = c_1 \psi_1 + c_2 \psi_2, \qquad \psi' = c'_1 \psi_1 + c'_2 \psi_2,
\end{equation}
where $c_1, c_2$ and $c'_1, c'_2$ are the two pair of coefficients obtained previously.
Here $\psi$ and $\psi'$ are the wave functions corresponding to two energy states $ E_0 +E^{(1)}_{+}$ and $E_{0} +E^{(1)}_{-}$, respectively, where $E_0=0$ in our case, and $E^{(1)}_{+}$ and $E^{(1)}_{-}$ are the modifications of the energy due to the effect of the perturbation. From the above equation we obtain
\begin{equation}
\psi_1 =\frac{c'_2 \psi -c_2 \psi'}{c_1 c'_2 -c'_1 c_2}.
\end{equation}

After reintroducing the time factor, we obtain the time dependent wave functions as
\bea
\Psi_1 &=& \frac{e^{-\frac{i}{\hbar}E_0\tau}}{c_1 c'_2 -c'_1 c_2} \Bigg[c'_2 \psi e^{-\frac{i}{\hbar}E^{(1)}_{+}\tau} -c_2 \psi' e^{-\frac{i}{\hbar}E^{(1)}_{-}\tau}\Bigg]=\nonumber\\
&&\frac{1}{c_1 c'_2 -c'_1 c_2} \Bigg[c'_2 \psi e^{-\frac{i}{\hbar}E^{(1)}_{+}\tau} -c_2 \psi' e^{-\frac{i}{\hbar}E^{(1)}_{-}\tau}\Bigg].\nonumber\\
\eea
Note that $\Psi_1 = \psi_1$ at $\tau =0$. Then we use $\psi_1, \psi_2$ to represent $\psi, \psi'$, and hence $\Psi_1$ becomes the linear combination of the wave functions $\psi_1, \psi_2$, with the combination coefficients time dependent. The absolute value of the coefficient multiplying $\psi_2$ and integrating over the volume $V$, is (after squaring) the transition probability $w_{21}$.   Therefore we have
\bea
w_{21} &=& \frac{1}{c_1c'_2 -c'_1 c_2} \Bigg[ c_1 c'_2 S_{12} e^{-i\frac{E^{(1)}_{+}}{\hbar} \tau} +c'_1 c_2 S_{12} e^{-i\frac{E^{(1)}_{-}}{\hbar} \tau} \nonumber\\
&&+c_2 c'_2 S_{22}\left(e^{-i\frac{E^{(1)}_{+}}{\hbar} \tau}-e^{-i\frac{E^{(1)}_{-}}{\hbar} \tau}\right)
\Bigg] =\nonumber \\
& & \Bigg\vert \frac{Q_1}{Q_3} \left(e^{-i\frac{E^{(1)}_{+}}{\hbar} \tau} -e^{-i\frac{E^{(1)}_{-}}{\hbar} \tau}\right) +Q_2 e^{-i\frac{E^{(1)}_{-}}{\hbar} \tau}
\Bigg\vert ^2  \nonumber\\
&=& 2\left(\frac{Q_1^2}{Q_3^2}-\frac{Q_1 Q_2}{Q_3}\right)\Bigg[1-\cos\left(\frac{E^{(1)}_{+}-E^{(1)}_{-}}{\hbar}\tau \right)\Bigg]+\nonumber\\
&&Q_2^2,
\eea
where we have denoted
\be
Q_1 =c_1 c'_2 S_{12} +c_2 c'_2 S_{22}, Q_2 = S_{12}, Q_3 =c_1 c'_2 -c'_1 c_2.
\ee

With the use of the previous results, the transition probability in ordinary units is
\begin{equation}
w_{21}=\frac{0.0106435}{l^{10}} -\frac{0.00659062}{l^{10}} \cos \left(\frac{0.880802}{\hbar}\eta l^3 \tau\right).
\end{equation}
From the above  equation it follows that the transition probability is a cosine function of the time $\tau$. When $\tau$ is very small we have $w_{21}$ proportional to $\tau^2$. Since we know that the probability is smaller than one, we have the restriction $0.0106/l^{10} -0.0066/l^{10} \leq 1$, a condition which gives for the upper limit of $l$ the numerical value
\begin{equation}
l \geq 0.666 \rightarrow a (\tau) \leq 1.177.
\end{equation}

Therefore we get the upper limit of the size of the Universe for which our quantum model is applicable. And the lower limit of $y$ is $0.320 < 0.5$, and therefore we can approximate the lower limit in the integral as $l_1 =0$.

 \section{The second quantization of time}\label{sect5}

In standard quantum mechanics, time is not an operator. The energy of the system is the eigenvalue of the Hamiltonian, which in the case of the harmonic oscillator can be written in the form  $\hat{H} =\hbar \omega (a^{+}a +\frac{1}{2})$, where $a$ and $a^{+}$ are the creation and annihilation operators. On the other hand the deep connections existing between energy and time suggests us to find a way to define the creation and annihilation operators of 'time'. After these operators are found, we can get rid of the concept of time singularity at the beginning of the Universe, and we can properly define the distance between each pair of time slices. And, consequently, we can obtain the quantum frequency of the 3-space evolution, and therefore investigate from a quantum mechanical point of view the birth of the Universe.

In the previous Section, by using the mathematical formalism of the $f(R,T)$ gravity theory we have defined a 'time' variable in the WDW equation, and thus we have transformed it into a Schr\"{o}dinger-Wheeler-de Witt equation. The time $\tau$ we have defined earlier is based on the relation $P_{\tau} =-\frac{2}{6a^3} P_D P_M$, and therefore in our analysis we have assumed that the idea of time in the $f(R,T)$ gravity theory is related to the field variables  $D$ and $M$. In the following we want to define the creation/annihilation operators based on the term $3kaDM -\frac{2}{6a^3}P_D P_M$ in the WDW equation (\ref{H}). The procedure goes as follows.

We assume the existence of the classical and quantum analogy for the $f(R,T)$ gravity model, which allows us to turn the classical Poisson brackets into quantum commutators, $\left\{ ...\right\}\rightarrow \left[...\right]$. Thus we {\it postulate} the following commutation relations
\begin{equation}
\left[\hat{D},\hat{P_D}\right] = \left[\hat{M},\hat{P_M}\right] = i\hbar.
\end{equation}

Since
\begin{equation}
\left(\hat{D}+i\hat{M}\right)\left(\hat{P_D} -i\hat{P_M}\right) =\hat{D} \hat{P_D} -i\hat{D} \hat{P_M} +i \hat{M} \hat{P_D} +\hat{M} \hat{P_M},
\ee
\be
\left(\hat{P_D} -i\hat{P_M}\right)\left(\hat{D}+i\hat{M}\right) =\hat{P_D} \hat{D} -i\hat{P_M} \hat{D} +i \hat{P_D} \hat{M} +\hat{P_M} \hat{M},
\end{equation}
we have
\begin{equation}
\left(\hat{D}+i\hat{M}\right)\left(\hat{P_D} -i\hat{P_M}\right) - \left(\hat{P_D} -i\hat{P_M}\right)\left(\hat{D}+i\hat{M}\right) =2i\hbar.
\end{equation}
In the following we denote
\begin{equation}
\hat{v} = \hat{D}+i\hat{M},\qquad \hat{P_v} = \hat{P_D} -i\hat{P_M}.
\end{equation}

Hence, by using the mathematical identities $\left[\hat{P_D} ,\hat{M}\right] = \left[\hat{P_M} ,\hat{D}\right]=0$, $\hat{D}=\hat{D}^{*} ,\hat{M}=\hat{M}^{*},\hat{P_D}=\hat{P_D}^{*}, \hat{P_M}=\hat{P_M}^{*}$, where * denotes the complex conjugate, we obtain the commutation relations
\begin{equation}
[\hat{v}, \hat{P_v}] =2i\hbar,
\end{equation}
\bea
&&\hat{P_v}^{*} \hat{P_v}^{*} -\hat{P_v} \hat{P_v} =\left(\hat{P_D }+i\hat{P_M}\right)\left(\hat{P_D }+i\hat{P_M}\right) -\nonumber\\
&&\left(\hat{P_D }-i\hat{P_M}\right)\left(\hat{P_D} -i\hat{P_M}\right) =4i \hat{P_D} \hat{P_M},
\eea
\bea
\hat{v}^{*}\hat{v}^{*} -\hat{v}\hat{v} &=&\left(\hat{D}-i\hat{M}\right)\left(\hat{D}-i\hat{M}\right) - \left(\hat{D}+i\hat{M}\right)\times \nonumber\\
 &&\left(\hat{D}+i\hat{M}\right)= -4i \hat{M}\hat{D},
\eea
and
\bea
&&\hat{v}\hat{v} - \hat{v}^{*}\hat{v}^{*} + \hat{P_v} \hat{P_v} -\hat{P_v}^{*}\hat{P_v}^{*} =\left(\hat{v}-i\hat{P_v}\right)\left(\hat{v}+i\hat{P_v}\right) -\nonumber\\
&&\left(\hat{v}^{*} +i \hat{P_v}^{*}\right)\left(\hat{v}^{*} -i\hat{P_v} ^{*}\right),
\eea
respectively. Also
\be
\left[\hat{v}^{*},\hat{P_v}^{*}\right]=\left(2i\hbar\right)^{*}.
\ee

Let's assume now that there are {\it two directions of time}, defined as
\begin{equation}
\hat{\tau_1} =\hat{v}-i\hat{P_v },\qquad \hat{\tau_2} = \hat{v} +i \hat{P_v}.
\end{equation}

Then we have
\begin{equation}
\left[\hat{\tau_1} ,\hat{\tau_2}\right] =\left[\hat{v}- i\hat{P_v} , \hat{v}+ i\hat{P_v}\right] = -4\hbar , \left[\hat{\tau_1},\hat{\tau_1}\right] =\left[\hat{\tau_2},\hat{\tau_2}\right] =0.
\end{equation}

As a next step in our analysis we interpret $\hat{\tau_1}$ as a creation operator $\hat{\tau}^{+}$ and $\hat{\tau _2}$ as an annihilate operator $\hat{\tau}$.  Thus we further obtain
\bea
&&\hat{v}\hat{v} - \hat{v^{*}}\hat{v^{*}} + \hat{P_v}\hat{ P_v} -\hat{P_v^{*}}\hat{P_v^{*}} =\left(\hat{v}-i\hat{P_v}\right)\left(\hat{v}+i\hat{P_v}\right) -\nonumber\\
&&\left(\hat{v}^{*} +i\hat{ P_v}^{*} \right)\left(\hat{v^{*}} -i\hat{P_v ^{*}}\right)
=  \hat{\tau}^{+} \hat{\tau} -\left(\hat{\tau}^{+} \hat{\tau}\right)^{*}
= \nonumber\\
 &&\hat{N_{\tau}} - \hat{N^{*}_{\tau}} =
2i {\rm Im} \hat{N_{\tau}} =2i \hat{N}_{obs}.
\eea

Here we define the complex time "number" operator
\begin{equation}
\hat{N_{\tau}} \equiv \hat{\tau}^{+}\hat{\tau}
\end{equation}

The above relation shows that the time "number" operator, defined as
\begin{equation}
\hat{N}_{obs} \equiv {\rm Im} \hat{N_{\tau}},
\end{equation}
is an observable in quantum mechanics. Although the complex time has two directions, there is only one real time observable that can be measured in the experiments.

Now let's consider the case of our WDW equation, where
\begin{equation}
-\frac{2}{6a^3}P_D P_M + 3kaDM = \frac{2}{6a^3} \left(-\hat{P_D} \hat{P_M} +9ka^4 \hat{D}\hat{M}\right).
\end{equation}

In the following we discuss again the specific cosmological model $f(R,T)=R+\theta RT$. We rescale $\hat{v}$ as
\begin{equation}
\hat{v} = \sqrt{3a^2 k} \hat{ v}.
\end{equation}
 Thus we obtain the corresponding representation of the time-related terms in the WDW equation as
\bea
&&-\frac{2}{6a^3}\hat{P_D} \hat{P_M} + 3ka\hat{D}\hat{M} =\nonumber\\
&&\frac{-i}{12a^3}  \left(\hat{v}\hat{v} -\hat{v^{*}}\hat{v^{*}} +\hat{P_v} \hat{P_v} -\hat{P_v^{*}} \hat{P_v^{*}}\right) =\nonumber\\
&&\frac{-i}{12a^3}\Bigg[\hat{\tau}^{+}\hat{\tau} -\left(\hat{\tau}^{+}\hat{\tau}\right)^{*}\Bigg] =\frac{-1}{6a^3} {\rm Im} \hat{N}_{obs}.
\eea

From this relation it follows that the coupling between the gravitational field and the matter field gives us a way to measure the quantum time number of a given Universe. With the use of the above relations the WDW equation becomes
\begin{equation}\label{fin}
\Bigg[-2aP_a P_A +2P_A ^2 A -18ka^4 A -\frac{3a^6}{2}M \Bigg]\Psi = \hat{N}_{obs} \Psi.
\end{equation}

Eq.~(\ref{fin}) gives us the possibility of further constructing a wave function with $N$ quanta of time, which has the property
\begin{equation}
\hat{N}_{obs} \Psi_{N} = N \Psi_{N}
\end{equation}
Therefore the WDW equation becomes
\begin{equation}\label{fin1}
\Bigg[-2aP_a P_A +2P_A ^2 A -18ka^4 A -\frac{3a^6}{2}M \Bigg]\Psi_{N} = N\Psi_{N}.
\end{equation}

Eq.~(\ref{fin1}) may give us a clear picture of the evolution of the Universe when we use the Arnowitz-Deser-Misner (ADM) formalism, since the dynamics of system will change with the variation of the time quanta $N$. Therefore when we are at different time moments, we shall have different Schr\"{o}dinger equations to describe the local dynamics of the Universe.

Of course, we can also define the {\it observable time vacuum} (the beginning of the Universe) as
\begin{equation}
\tau \vert 0_{\tau}\rangle = \tau ^{*} \vert 0_{\tau^{*}}\rangle =0 , \qquad \vert 0_{obs} \rangle = \vert 0_{\tau} \rangle \vert 0_{\tau^{*}}\rangle
\end{equation}

In the present Section we have tried to introduce the Fock space of the quantum time variable $\tau$. If we define the time creation/annihilation operators in the Fock space, then we can get rid of the problem of time singularity at the beginning of the Universe, and we may have a deeper understanding of the discrete nature of time. But this will also require to transform our wave function into the quantum occupation number picture. However, we must note that this technique works only when the canonical momentum and its corresponding canonical position are not coupling with each other. One of the difficulties of the canonical quantization of gravity is that it cannot define the Hilbert space. The second quantization of time procedure may provide a prospective way to think about the problems of quantum gravity in the Fock space, instead of in the  Hilbert space.

\section{Discussions and final remarks}\label{sect6}

Quantum cosmology offers a lange number of challenges,  but also interesting insights into the fundamental nature of the space-time. The  physical problems related to the birth and very early evolution of the Universe  might be better understood by using the mathematical formalism of quantum theory, including symmetries, discrete structures, or semi-classical features extracted from a generally covariant, and highly interacting quantum theory.

In this paper we have investigated the quantum cosmology of the $f(R,T)$ gravity theory, a modified gravity theory in which the gravitational action is an arbitrary function of the Ricci scalar and of the trace of the energy-momentum tensor. In the present paper we have considered that the classical evolution of the Universe takes place in the background Friedmann-Robertson-Walker geometry, which we are using systematically to investigate the quantum properties of the early Universe. As a starting point in our analysis we have introduced the Hamiltonian formulation of the theory, which is constructed systematically from the action given by Eq.~(\ref{L}). In the action we have introduced two Lagrange multipliers $\lambda $ and $\mu$, with the first imposing the (purely geometric) definition of the Ricci scalar, while the second one imposes the trace constraint of the $f(R,T)$ gravity theory. This constraint goes beyond a simple definition of the trace of the energy-momentum tensor, since it allows the investigation of the deep connection between matter and geometry at a more general level than the one that follows from the simple thermodynamic definition of $T$. From the cosmological gravitational action one can obtain the gravitational cosmological Hamiltonian, which, by canonical quantization, leads immediately to the general form of the Wheeler-de Witt equation, describing the evolution of the wave function of the quantum Universe. In order to obtain some physical insights in the quantum properties of the Universe we consider a simple extension of the standard general relativity, in which the gravitational Lagrangian is a "deformation" of the form $\theta RT$ of the general relativistic Lagrange function $R$. We have investigated in detail the properties of this quantum cosmological model. Its most interesting feature is the possibility of the definition of a quantum time, and of an associated canonical momentum operator. This leads to the reformulation of the Wheeler--de Witt equation as a Schr\"{o}dinger type equation. We have studied in detail the mathematical properties of this equation, by using a perturbative approach, in which the small perturbation is proportional to the energy of the system, in the framework of a twofold degenerate quantum system. The probability of transition between states is also obtained. As a theoretical possibility we have also discussed very briefly the second quantization of time, which leads to the interesting possibility of the extension of the Hilbert space of the canonical quantization method to the Fock space description of quantum phenomena in the very early Universe.

The initial state of the Universe is essentially unknown. That's why the possibility that the initial geometry of the Universe was not an isotropic and homogeneous, Friedmann-Robertson-Walker type one, cannot be rejected {\it a priori}. This raises the interesting question of the applicability of the formalism developed in the present to describe the quantum cosmology of $f(R,T)$ gravity to more general geometries. In particular, in the following we briefly consider the quantum cosmology of $f(R,T)$ gravity in the anisotropic Bianchi type I geometry, with the metric given by
\begin{equation}
ds^2  = -N^2(t)dt^2 +a_1^2(t)dx^2 +a^2_2 (t) dy^2 +a_3^2 (t)  dz^2,
\end{equation}
where $a_i$, $i=1,2,3$ are the directional scale factors. For the Bianchi type I geometry the scalar curvature is obtained as
\bea
R &=& \frac{2}{N^2}\Bigg[\left(\frac{\ddot{a_1}}{a_1}+\frac{\ddot{a_2}}{a_2}+\frac{\ddot{a_3}}{a_3}\right)+\left(\frac{\dot{a_1}\dot{a_2}}{a_1 a_2}+\frac{\dot{a_1}\dot{a_3}}{a_1 a_3}+\frac{\dot{a_2}\dot{a_3}}{a_2 a_3}\right) \nonumber\\
 &&-\frac{2\dot{N}}{N}\left(\frac{\dot{a_1}}{a_1}+\frac{\dot{a_2}}{a_2}+\frac{\dot{a_3}}{a_3}\right)\Bigg].
\eea
For the Bianchi type I geometry the gravitational action reads
\bea
S_g &=& \int dt \Bigg\{Na_1 a_2 a_3 f(R,T) -\lambda \left[R - \frac{2}{N^2}(\cdots)\right] -\nonumber\\
&&\mu \left[\frac{1}{2}M - \cdots\right] \Bigg\},
\eea
while the gravitational Lagrangian can be obtained as
\bea
\mathcal{L}_g &=& -\frac{2\widetilde{\lambda}}{N}\left(\dot{a_1}\dot{a_2}a_3 +\dot{a_1}\dot{a_3}a_2 + \dot{a_2}\dot{a_3}a_1\right) -\nonumber\\
 &&\frac{2}{N}\dot{\widetilde{\lambda}}\frac{d}{dt}\left(a_1 a_2 a_3)+ \frac{a_1 a_3}{a_2} +\frac{a_1 a_2}{a_3}\right) -Na_1 a_2 a_3 V.\nonumber\\
\eea
Note that all the definitions of $\widetilde{\lambda}, \widetilde{\mu}, V$ remind unchanged. In the following we
introduce a new variable $W = (a_1 a_2 a_3)^{\frac{1}{3}}$. Then we can easily obtain $\dot{W}=\frac{d}{dt}(a_1 a_2 a_3)/3V^2$, and
$3W\dot{W} = \frac{d}{dt}(a_1 a_2 a_3)$, respectively, as well as the relation
\bea
W \dot{W}^2 &=& \frac{W^3}{9} \sum _{i=1}^3{\left(\frac{\dot{a_i}}{a_i}\right)^2} +
\frac{2}{9}\left(\dot{a_1}\dot{a_2}a_3 +\dot{a_1}\dot{a_3}a_2 + \dot{a_2}\dot{a_3}a_1\right).\nonumber\\
\eea

Hence the gravitational Lagrangian of the $f(R,T)$ gravity theory in a Bianchi type I geometry can be represented as
\bea\label{167}
\mathcal{L}_{g} &=& -\frac{9\widetilde{\lambda}}{N} W\dot{W}^2 + \frac{\widetilde{\lambda}}{N}W^3 \sum _{i=1}^3{\left(\frac{\dot{a_i}}{a_i}\right)^2}-\nonumber\\
&& \frac{6}{N}\dot{\widetilde{\lambda}}W^2 \dot{W}  -NW^3 V.
\eea

The Hamiltonian corresponding to the Lagrangian (\ref{167}) now reads
\begin{equation}
H_g =\dot{W}P_W + \dot{a_1} P_{a_1} +\dot{a_2} P_{a_2}+\dot{a_3} P_{a_3} +\cdots -\mathcal{L}_g,
\end{equation}
and we obtain the canonical momenta corresponding to the variables  $\left(W, a_1, a_2, a_3\right)$ as
\bea
P_W &=& - \frac{18\widetilde{\lambda}}{N}W\dot{W} - \frac{6}{N}\dot{\widetilde{\lambda}}W^2,
P_{a_1} = 2\frac{\widetilde{\lambda}}{N}W^3\frac{\dot{a_1}}{a_1 ^2}, \nonumber\\
P_{a_2} &=& 2\frac{\widetilde{\lambda}}{N}W^3\frac{\dot{a_2}}{a_2 ^2},
P_{a_3} = 2\frac{\widetilde{\lambda}}{N}W^3\frac{\dot{a_3}}{a_3 ^2}.
\eea

For the other canonical momenta we obtain the following correspondence between the isotropic and anisotropic case,
\bea
P_q &=& \frac{-6}{N}a^2 \dot{a} (\cdots) \rightarrow P_q = -\frac{2}{N}\frac{d}{dt}(a_1 a_2 a_3) = \nonumber\\
&&-\frac{6}{N}W^2 \dot{W} (\cdots).
\eea
Hence the gravitational Hamiltonian becomes
\bea
H_g &=& -\frac{9\widetilde{\lambda}}{N}W\dot{W}^2 +\frac{2\widetilde{\lambda}}{N} W^3 \sum _{i=1}^3{\left(\frac{\dot{a_i}}{a_i}\right)^2}   +\nonumber\\
&& NW^3 V -\frac{6}{N}W^2 \dot{W} [\cdots],
\eea
where $[\cdots]$ is represented by
\begin{equation}
-\frac{6}{N}W^2 \dot{W} [\cdots] = -\frac{N}{6W^2} P_W \left(P_A +\frac{P_F}{3}\right) +\frac{18}{N} \widetilde{\lambda} W \dot{W}^2.
\end{equation}
Since
\begin{equation}
\frac{2\widetilde{\lambda}}{N}W^3 \frac{\dot{a_i}^2}{a_i ^2} = \frac{Na_i ^2}{2W^3 \widetilde{\lambda}}P_{a_i}^2,i=1,2,3,
\end{equation}
for the Hamiltonian we obtain
\bea\label{175}
H_g &=& \frac{9\widetilde{\lambda}}{N}W\dot{W}^2 - \frac{N}{6W^2}P_W \left(P_A +\frac{P_F}{3}\right) + \nonumber\\
&&\frac{N}{2W^3 \widetilde{\lambda}}\sum _{i=1}^3{a_i^2 P_{a_i}^2}  +Na_1 a_2 a_3 V.
\eea
The anisotropic cosmological Hamiltonian (\ref{175}) of the $f(R,T)$ gravity theory is very similar to the one obtained in the isotropic case.

When considering the $f(R,T) = F^0 (R) + \theta RT$ model in the case of $R \rightarrow \infty$, we obtain
\begin{equation}
\dot{D} = -2 \frac{N}{4W^3}P_M , P_M = \frac{1}{2}\frac{6}{N}W^2 \dot{W} D,
\end{equation}
and hence  we immediately arrive at $\dot{D} = - (3/2) \left(\dot{W}/W\right) D$, and $D = \delta _0/W^{3/2}$,
respectively, where $\delta _0$ is a constant. Since
\be
P_D = -i\frac{\partial}{\partial D} = -i \frac{dW}{dD}\frac{\partial}{\partial W} = i\frac{2W^{5/2}}{3\delta _0} \frac{\partial}{\partial W},
\ee
then we have
\begin{equation}
-\frac{2}{4W^3}P_D P_M =-\frac{6\dot{W}}{4W}P_D D = -i\dot{W} \frac{\partial}{\partial W}.
\end{equation}
\\
By taking into account that
\begin{equation}
\frac{N}{2W^3 \widetilde{\lambda}} a_i ^2 P_{a_i}^2 = \dot{a_i}P_{a_i} = -i a_i \frac{\partial}{\partial a_i}, i= 1,2,3,
\end{equation}
\\
we obtain the time canonical momentum
\bea
P_{\tau} &=& -i\frac{\partial}{\partial \tau} =-i[ \frac{dW}{d\tau} \frac{\partial}{\partial W}+\sum _{i=1}^3\frac{da_i}{d\tau}\frac{\partial}{\partial a_i}=\nonumber\\
&&= -\frac{2}{4W^3}P_D P_M + \frac{1}{2W^3 \widetilde{\lambda}}\sum _{i=1}^3{a_i^2 P_{a_i}^2}.
\eea

Finally,  the transformation
\begin{equation}
-\frac{2}{4W^3}P_D P_M + \frac{1}{2W^3 \widetilde{\lambda}}\sum _{i=1}^3{a_i^2 P_{a_i}^2} \rightarrow P_{\tau},
\end{equation}
will allow us the introduce the time dependence of the Wheeler-de Witt equation for anisotropic Bianchi type I geometries in the $f(R,T)$ gravity theory.
Note that here we have assumed that $a_1(t)$, $a_2(t)$, $a_3(t)$, $W(t)$ are independent variables.  Hence we can safely conjecture that in the anisotropic case we can still introduce a cosmological quantum time in the Wheeler- de Witt equation of $f(R,T)$. On a qualitative level the overall results of the anisotropic case will be very similar to the ones obtained for isotropic and homogeneous geometries.

In the present paper we have introduced some basic theoretical tools that could be used for the investigation of the quantum properties of the gravitational interaction, and of the evolution and origin of the very early Universe, in which the complex interaction of geometry and matter  give birth to time, entropy, and irreversibility.

\section*{Acknowledgments}

We would like to thank to the anonymous referee for comments and suggestions that helped us to significantly improve the manuscript. T. H. would like to thank the Yat Sen School of the Sun Yat Sen University in Guangzhou, P. R. China,  for the kind hospitality offered during the preparation of this work.

\appendix

\section{The variation of the gravitational action in the $f(R,T)$ gravity theory}\label{App1}

By varying the gravitational action (\ref{act}) of the $f(R,T)$ theory with respect to the metric tensor we obtain first

\bea
\delta S&=&\frac{1}{16\pi}\int [f_R(R,T)\delta R+f_T (R,T)\frac{\delta T}{\delta g^{\mu\nu}}\delta g^{\mu\nu} -\nonumber\\
&&\frac{1}{2}g_{\mu\nu} f(R,T) \delta g^{\mu\nu} +
16\pi \frac{1}{\sqrt{-g}}\frac{\delta (\sqrt{-g}L_m)}{\delta g^{\mu\nu}}]\sqrt{-g} d^4 x. \nonumber\\
\eea

For the variation of the Ricci scalar we obtain
\begin{equation}
\delta R = \delta(g^{\mu\nu}R_{\mu\nu}) = R_{\mu\nu}\delta g^{\mu\nu} +g^{\mu\nu} (\nabla _{\lambda}\delta \Gamma^{\lambda}_{\mu\nu}-\nabla_{\nu} \delta \Gamma ^{\lambda}_{\mu\lambda}).
\end{equation}
Since
\begin{equation}
\delta \Gamma ^{\lambda}_{\mu\nu}=\frac{1}{2}g^{\lambda\sigma}(\nabla_{\mu}\delta g_{\nu\alpha}+\nabla_{\nu}\delta g_{\alpha\mu}-\nabla_{\alpha}\delta g_{\mu\nu}),
\end{equation}
we finally obtain
\begin{equation}
\delta R =R_{\mu\nu} \delta g^{\mu\nu} +g_{\mu\nu} \Box \delta g^{\mu\nu} -\nabla_{\mu}\nabla_{\nu}\delta g^{\mu\nu}.
\end{equation}

Therefore we obtain the variation of the action as
\bea
\delta S &=&\frac{1}{16\pi} \int \Bigg[f_R (R,T) R_{\mu\nu} \delta g^{\mu\nu} +f_R (R,T) g_{\mu\nu} \Box \delta g^{\mu\nu} -\nonumber\\
&& f_R (R,T) \nabla_{\mu}\nabla_{\nu}\delta g^{\mu\nu}+f_T (R,T) \frac{\delta(g^{\alpha\beta}T_{\alpha\beta})}{\delta g^{\mu\nu}} \delta g^{\mu\nu}-\nonumber\\ &&\frac{1}{2}g_{\mu\nu} f(R,T) \delta g^{\mu\nu}+16\pi \frac{1}{\sqrt{-g}}\frac{\delta (\sqrt{-g}L_m)}{\delta g^{\mu\nu}}\Bigg]\sqrt{-g}d^4 x. \nonumber\\
\eea

We define the variation of $T$ with respect to the metric tensor as
\begin{equation}
\frac{\delta (g^{\alpha\beta}T_{\alpha\beta})}{\delta g^{\mu\nu}} =T_{\mu\nu}+\Theta _{\mu\nu},
\end{equation}
where
\begin{equation}
\Theta _{\mu\nu} \equiv g^{\alpha\beta}\frac{\delta T_{\alpha\beta}}{\delta g^{\mu\nu}}
\end{equation}

After partially integrating the second and the third term in the variation of the action, we obtain the field equations for our model as
\bea
&&f_R(R,T) R_{\mu\nu} -\frac{1}{2} f(R,T) g_{\mu\nu} +(g_{\mu\nu}\Box -\nabla_{\mu}\nabla_{\nu}) f_R{R,T} \nonumber\\
&&= -8\pi T_{\mu\nu} -f_T (R,T) T_{\mu\nu} -f_T (R,T) \Theta _{\mu\nu}.
\eea

\section{The energy balance equations in $f(R,T)$ gravity}\label{App2}

With the use of the  mathematical identity
\begin{equation}
\nabla^{\mu}[f_R R_{\mu\nu} -\frac{1}{2}fg_{\mu\nu} +(g_{\mu\nu}\Box - \nabla_{\mu}\nabla_{\nu}) f_R]\equiv -\frac{1}{2}f_T \nabla^{\mu} T g_{\mu\nu},
\end{equation}
by taking the covariant divergence of the field equations (\ref{f1})  we obtain first
\bea
-\frac{1}{2}f_T \nabla^{\mu} T g_{\mu\nu} &=& 8\pi \nabla^{\mu} T_{\mu\nu} +\nabla ^{\mu} f_T  T_{\mu\nu}+ f_T \nabla^{\mu}T_{\mu\nu} -\nonumber\\
&&\nabla^{\mu} f_T pg_{\mu\nu} -f_T \nabla ^{\mu}p g_{\mu\nu}
\eea

Therefore we have
\begin{equation}
\nabla^{\mu}T_{\mu\nu} =\frac{f_T}{8\pi +f_T}[(pg_{\mu\nu}-T_{\mu\nu})\nabla^{\mu}Inf_T +(\nabla^{\mu}p -\frac{1}{2}\nabla^{\mu}T)g_{\mu\nu}].
\end{equation}

In the following we  adopt for the energy-momentum tensor the perfect fluid form
\begin{equation}
T_{\mu\nu} =(\epsilon +p )U_{\mu}U_{\nu} +pg_{\mu\nu}.
\end{equation}
In the comoving reference frame $U_{\mu}=(N(t),0,0,0)$, $U^{\mu}=(-1/N(t),0,0,0)$, $L_m =p$. and therefore $\Theta_{\mu\nu}=-2T_{\mu\nu} +pg_{\mu\nu}$. Hence we find
\bea
\nabla^{\mu} T_{\mu\nu} &=& \left(\nabla^{\mu} \epsilon +\nabla ^{\mu}p\right)U_{\mu}U_{\nu} +(\epsilon +p )\nabla ^{\mu} U_{\mu}U_{\nu} +\nonumber\\
&&(\epsilon +p)U_{\mu}\nabla ^{\mu}U_{\nu} +\nabla ^{\mu}p g_{\mu\nu}.
\eea
And since
\begin{equation}
\nabla^{\mu}T_{\mu\nu} = \frac{f_T}{8\pi + f_T}[-(\epsilon +p)U_{\mu}U_{\nu}\nabla ^{\mu}Inf_T +\frac{1}{2}\nabla ^{\mu}(\epsilon -p)g_{\mu\nu}],
\end{equation}
we obtain
\bea
&&\left(\nabla^{\mu} \epsilon +\nabla ^{\mu}p\right)U_{\mu}U_{\nu} +(\epsilon +p )\nabla ^{\mu} U_{\mu}U_{\nu} +\nonumber\\
&&(\epsilon +p)U_{\mu}\nabla ^{\mu}U_{\nu} +
\nabla ^{\mu}p g_{\mu\nu} =\nonumber\\
&&\frac{f_T}{8\pi + f_T}\Bigg[-(\epsilon +p)U_{\mu}U_{\nu}\nabla ^{\mu}\ln f_T +\frac{1}{2}\nabla ^{\mu}(\epsilon -p)g_{\mu\nu}\Bigg].\nonumber\\
\eea

Multiplying $U^{\nu}$, and by taking into account the geodesic equation $U^{nu}\nabla ^{\mu}U_{\nu}=0$, we have
\bea
&&U_{\mu}\nabla^{\mu}\epsilon +(\epsilon +p)\nabla ^{\mu}U_{\mu} =\nonumber\\
&&
-\frac{f_T}{8\pi +f_T } \Bigg[(\epsilon +p)U_{\mu} \nabla^{\mu}\ln f_T +\frac{1}{2}U_{\mu}\nabla^{\mu}(\epsilon -p)\Bigg].\nonumber\\
\eea

\section{The components of the Ricci tensor and the Ricci scalar in the FRW geometry}\label{App3}

By simple calculations we obtain the components of the Ricci tensor for the FRW metric (\ref{FRW}) as
\begin{equation}
R_{00} = -3 \frac{\ddot{a}}{a} + 3 \frac{\dot{N}\dot{a}}{Na},
\end{equation}
\begin{equation}
R_{ii}=\frac{g_{ii}}{N^2}\Bigg[\frac{\ddot{a}}{a}+2(\frac{\dot{a}}{a})^2 +\frac{2kN^2}{a^2}-\frac{\dot{N}\dot{a}}{Na}\Bigg], i=1,2,3.
\end{equation}
Therefore for the Ricci scalar we obtain
\begin{equation}
R = \frac{6}{N^2}\left[\frac{\ddot{a}}{a}+\left(\frac{\dot{a}}{a}\right)^2 +\frac{kN^2}{a^2}-\frac{\dot{N}\dot{a}}{Na}\right].
\end{equation}

\section{The canonical momenta of the cosmological action in $f(R,T)$ gravity}\label{App4}

The canonical  momenta associated to the cosmological action Eq.~(\ref{L}) of the $f(R,T)$ gravity are given by
\begin{equation}
P_a = -2\frac{6}{N}a \dot{a} \widetilde{\lambda} -\frac{6}{N} a^2 \dot{\widetilde{\lambda}},
 P_A =-\frac{6}{N} a^2 \dot{a} \left(1-\frac{B}{\mathcal{A}}\right),
\end{equation}
\begin{equation}
P_B =-\frac{6}{N} a^2 \dot{a} \left(-\frac{\mathcal{Z}}{\mathcal{A}} +\frac{3B\mathcal{Z}}{\mathcal{A}^2}\right),
P_C =-\frac{6}{N} a^2 \dot{a} \left(\frac{BR}{\mathcal{A}}\right),
\end{equation}
\begin{equation}
P_D =-\frac{6}{N} a^2 \dot{a}\left[-\frac{B(T-4p)}{\mathcal{A}}-\frac{B\mathcal{Z}R}{\mathcal{A}^2}\right],
\ee
\be
P_E = -\frac{6}{N} a^2 \dot{a} \left[\frac{B\mathcal{Z}(T-4p)}{\mathcal{A}^2}\right],
P_F = -\frac{6}{N} a^2 \dot{a} \left( \frac{3B}{\mathcal{A}}\right),
\end{equation}
\begin{equation}
P_G = -\frac{6}{N} a^2 \dot{a} \left(-\frac{3B\mathcal{Z}}{\mathcal{A}^2}\right),
P_R = -\frac{6}{N} a^2 \dot{a} \left[\frac{BC}{\mathcal{A}}-\frac{B\mathcal{Z}D}{\mathcal{A}^2}\right],
\end{equation}
\begin{equation}
P_T =  -\frac{6}{N} a^2 \dot{a} \left(-\frac{BD}{\mathcal{A}}+\frac{B\mathcal{Z}E}{\mathcal{A}^2}\right),
\ee
\be
P_p =  -\frac{6}{N} a^2 \dot{a} \left(\frac{4DB}{\mathcal{A}}-\frac{4B\mathcal{Z}E}{\mathcal{A}^2}\right).
\end{equation}

\end{document}